\documentclass[12pt]{article}

\usepackage[a4paper]{geometry}

\usepackage{setspace}
 
\usepackage{mathptmx}
\usepackage{newtxtext,newtxmath}

\usepackage{amsmath}

\usepackage{bm}

\usepackage{graphics}
\usepackage{epsfig}
\usepackage{colortbl}
\usepackage{color}
\usepackage{nicefrac}
\usepackage{mathrsfs}
\usepackage{bbm}

\usepackage{subfig}
\usepackage{booktabs}

\usepackage[breaklinks=true,colorlinks=true,linkcolor=blue,urlcolor=blue,citecolor=blue]{hyperref}

\RequirePackage{color}
\usepackage{soul}

\begin{document}

\begin{titlepage}
	
	\vskip 1.5 cm
	
		\begin{center}
	\Large Partition function zeros for the Blume-Capel model on a complete graph
	\end{center}
	
	\vskip 1.5 cm
	\centerline{{\bf{Yulian Honchar}}$^{1, 2}$, {\bf{Mariana Krasnytska}}$^{1, 2, 3}$, 
		
        {\bf Bertrand Berche}$^{2,4}$,
		{\bf{Yurij Holovatch}}$^{1,2,5,6}$,
        \fbox{{\bf{Ralph Kenna}}}$^{2, 5}$
		}	\vskip 0.5 cm
	\begin{center}	
		$^1$
		Institute of Condensed Matter Physics, National Academy of Sciences of Ukraine, UA -- 79011 Lviv, Ukraine
	\\ \vspace{0.5cm}
		$^2$ 
		${\mathbb L}^4$ Collaboration \& Doctoral College for the Statistical Physics of Complex Systems, Leipzig-Lorraine-Lviv-Coventry, Europe
		\\ \vspace{0.5cm}
        $^3$
         Haiqu, Inc., Shevchenka St, 120G, Lviv, Ukraine, 79039 	\\ \vspace{0.5cm}
         $^4$
      Laboratoire de Physique et Chimie Th\'eoriques, Universit\'e de Lorraine - CNRS, UMR 7019, Nancy, France	\\ \vspace{0.5cm}
		$^5$ 
		Centre for Fluid and Complex Systems, Coventry University, Coventry, CV1 5FB, United Kingdom
		\\ \vspace{0.5cm}
        $^6$ Complexity Science Hub, 1030 Vienna, Austria

	\end{center}

	\begin{abstract}
		In this paper we study finite-size effects in the Blume-Capel model through the analysis of the zeros of the partition function.   We consider a complete graph and make use of the behaviour of the partition function zeros to elucidate the crossover from effective to asymptotic properties.
        While in the thermodynamic limit the exact solution yields the asymptotic mean-field behaviour, for finite system sizes an effective critical behaviour is observed.        
        We show that even for large systems, the criticality is not asymptotic. We also present insights into how partition function zeros in different complex fields (temperature, magnetic field, crystal field) give different precision and provide us with different parts of the larger picture. This includes the differences between criticality and tricriticality as seen through the lens of Fisher, Lee-Yang, and Crystal Field zeros.
			
	\end{abstract}
	Keywords: Blume-Capel model, finite-size effects, partition function zeros.
\end{titlepage}

\clearpage

\section{{Introduction}}\label{0}
 The study of phase transitions and critical phenomena is a fundamental part of statistical physics. These processes, where systems undergo drastic changes in their behaviour due to variations in parameters like temperature, have been of interest for over a century. Among the many models developed to understand these phenomena, the Blume-Capel model has proven to be simple yet very versatile. First introduced in the 1960s, this model extends the well-known Ising model by including a single-ion anisotropy term \cite{blume_theory_1966, capel_possibility_1966}. The Blume-Capel model is particularly valuable because it exhibits a rich phase diagram, including first- and second-order phase transitions, and the tricritical point that connects them.

While the exact solution of the Blume-Capel model on a complete graph is well-known, finite-size effects, especially those concerning the model's critical behaviour, remain largely underexplored. In statistical physics, understanding finite-size effects is crucial, as real-world systems are always of finite size. These effects can significantly alter the observed critical properties and provide valuable insights into the system's behaviour as the its size grows. Moreover, finite-size scaling allows one to predict how systems behave as they approach infinite size, which is often a goal in theoretical studies.

To investigate the finite-size behaviour of the Blume-Capel model on a complete graph, this study utilizes the analysis of the partition function zeros, specifically zeros in complex magnetic field, temperature, and crystal field planes. This approach has been successfully applied to various models in statistical physics to identify critical points and characterize phase transitions. The distribution of these zeros provides information about the nature of the phase transition and the system's critical properties. Additionally, we will introduce an expanded function approximation that simplifies the calculations without sacrificing accuracy. This method, while not exact, is effective and offers a more computationally efficient way to study the model's behaviour for finite-size systems.

In Section \ref{sec_model} we remind the reader about the Blume-Capel model and point out its important limits. In Section 
\ref{sec_exact} we elaborate on the phase diagram of the Blume-Capel model in the thermodynamic limit. Section \ref{sec_method}
describes the methodology of the finite-size scaling and specifically the partition function zeros analysis that will be 
used in this paper. Section \ref{sec_zeros} gives the main results for the effective behaviour of the finite-size complete 
graph. Conclusions and outlook are given in Section \ref{sec_conclusions}.

\section{The model Hamiltonian and partition function} \label{sec_model}
\begin{figure}[h!]
    \centering
    \includegraphics[width=0.8\linewidth]{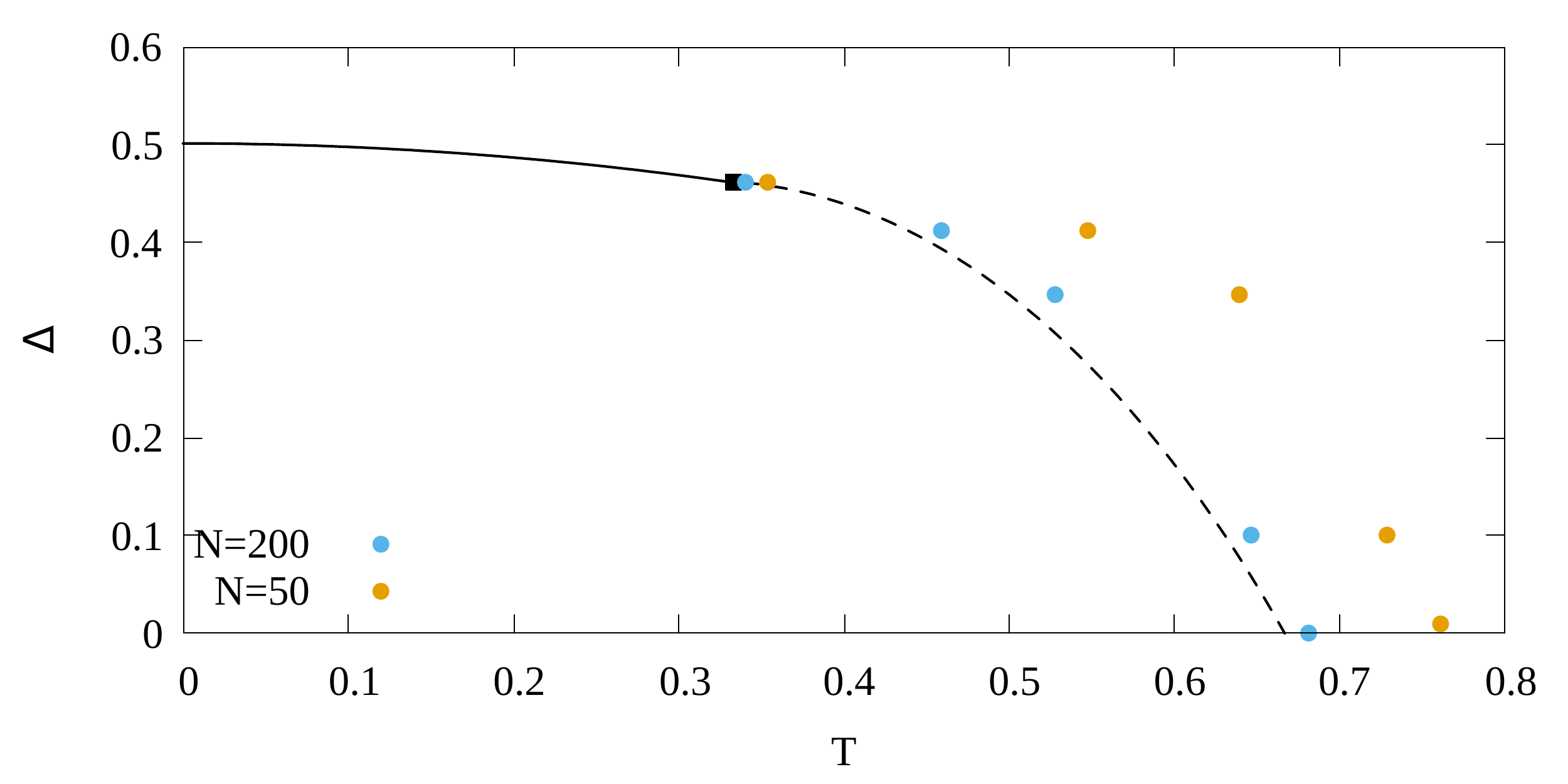}
    \caption{Phase diagram for the Blume-Capel model on a complete graph. The low temperature and the low crystal field sector is an ordered phase, while as the temperature rises, the system becomes disordered. The dashed line represents a line of critical points and a solid line is the line of the first-order phase transitions. The point at which they meet is the tricritical point, see Section \ref{sec_exact} for more details. Coloured points to the right of the curve denote the effective points of second-order phase transitions for finite-size systems
    of $N=50$ and $N=200$ nodes, correspondingly, obtained in Section \ref{sec_zeros}  by the partition function zeros analysis. As expected, the larger the number of nodes, the closer the data points to the expected asymptotic curve.  } 
    \label{fig:phase_full}
\end{figure}

In 1966 Alexander Blume introduced a variant of the Ising model, with the
physical motivation of studying magnetization in Uranium Oxide \cite{blume_theory_1966}. Later, Hans Capel modified it to demonstrate the possibility of first-order phase transitions in Ising systems of triplet ions with zero-field splitting \cite{capel_possibility_1966}. Here we consider Hamiltonian for the Blume-Capel model on a complete graph:
\begin{equation}\label{Hamiltonian}
{\cal  H}=-\frac{J}{2N}\sum_{l\neq m}S_lS_m + \Delta \, \sum _l S_l^2 - H\, \sum _l S_l \, , \hspace{1cm}S_l=-1,0,1,
\end{equation} 
where $\Delta$ and $H$ are crystal and external magnetic fields, the sums span all $N$ nodes of the graph and we fix the energy
scale by putting  $J=1$ in what follows below (the factor $1/N$ in the spin-spin coupling is required to restore extensivity of the energy when all spin pairs interact). This three-state spin model has an interesting phase landscape. In the absence of the ordering field $H$ the  Blume-Capel 
$\Delta - T$ plane becomes an important case study, see Fig.~\ref{fig:phase_full}. In particular, it features lines of the second- and first-order phase transitions, connected with the tricritical point. For the complete graph, the tricritical point is located at $\Delta_t = 2/3 \ln 2, \,T_t=1/3$, see also the next section \ref{sec_exact}.
The plot shows also the finite size results as described in more detail in Section \ref{sec_zeros}.

Let us revisit some of the defining limits of this model.
The $T-\Delta$ plane at $H=0$ is called the \textit{symmetric plane} \cite{lawrie_theory_1984} and there the Hamiltonian reads:
\begin{equation}
    {\cal  H}= -\frac{1}{2N} \sum_{l\neq m} S_l S_m + \Delta\sum_l S_l^2\, ,
\end{equation}
$N$ being the total number of sites.

\begin{itemize}
\item \textit{The limit $T \rightarrow 0$ for $\Delta \geq 0$:} In the $T \rightarrow 0$ limit all spins are frozen to one of three states $-1, 0$ or $1$. If they
are all $\pm 1$, then $\beta {\cal  H}= (-\frac{N-1}{2NT} + \frac{\Delta}{T})N$ (from now on we take $\beta=1/T$ with the Boltzmann constant $k_B=1$ for simplicity).
 In this circumstance, the system is magnetised and in the usual ordered phase.
If the spins are frozen to $0$ the system has zero magnetisation (although is in a type of ordered phase). The equal weight criterion gives that there is the first-order transition when the free energies are equal, i.e. when $1/2T - \Delta/T = 0$. This occurs at $\Delta=1/2$ (see the phase diagram in Fig. \ref{fig:phase_full}). The magnetised $(\pm 1)$ phase is dominant for $\Delta<1/2$ while the unmagnetised phase is dominant for $\Delta>1/2$. To see this, consider $\Delta = 1/2 + \varepsilon$, for which $\exp(-\beta {\cal H}) = \exp(-\varepsilon\Delta/T)$. If $\varepsilon>0$, the nonzero terms are washed away so that $S_i=0$ dominates and vice versa for $\varepsilon<0$.

\item \textit{The limit $\Delta=0$:} In this case we have a three-state Ising model, where all three spin states are equally possible. The critical point here is located at the temperature $T=2/3$.

\item \textit{The limit $\Delta \rightarrow - \infty$ at fixed $T$:} In this case $\exp(-\beta {\cal H}) = \exp(+\infty\sum S_i^2)$, so that only the values $S_i=\pm1$ survive. This is a pure (two-state) Ising model with critical temperature $T=1$.

\item \textit{The limit $T\rightarrow\infty$ (or $J\rightarrow0$) at fixed $\Delta/T$:} Here we have $N$ independent spins so that the partition function factorises into $N$ independent partition functions, one for each spin. I.e., $Z=z_i^N$, where $z_i=2e^{-\Delta/T}+1$. The equal weight condition applied to each of these gives a pseudo-transition at $\exp(\Delta/T)=1/2$ or $\Delta=T\ln2$.
\end{itemize}
To proceed obtaining an integral representation for
the partition function of the model (\ref{Hamiltonian}) we make use of
the equality $(\sum_{l}S_l)^2=\sum_{l}S_l^2 +\sum_{l\neq m}S_lS_m$ to factorize
the $N$-particle partition function:
\begin{eqnarray} \label{II.2}
Z_N(T,\Delta,H)= {\rm Sp}\, {\rm e}^{- {\cal H}/T} =
\prod_{l=1}^{N}\sum_{S_l=\pm 1,0}\exp\Big(\frac{
1}{2TN}(\sum_{l}S_l)^2-
\frac{1}{T} \sum _l S_l^2(\frac{1}{2N}+\Delta)+\frac{H}{T}\, \sum _l S_l\Big)\,.
  \end{eqnarray} 
Applying the Stratonovich-Hubbard transformation we express the
partition function in the form where the summation over $S_l$ can be
taken exactly. Finally, one arrives at the following integral representation
for the partition function of the Blume-Capel model on the complete graph:
\begin{eqnarray} \label{II.4}
Z_N (T,\Delta,H) = 
\int_{-\infty}^{+\infty} \exp\Big (\frac{-Nx^2}{2T}+N \ln(1+
2e^{-(\frac{1}{2N}+\Delta)/T}
\cosh[(x+H)/T])\Big)dx \, ,
  \end{eqnarray}
here and below we omit the 
prefactors irrelevant for the forthcoming analysis.
Since the integral representation expression (\ref{II.4}) for the partition function is exact, it can be analysed for any finite
value of $N$ as well as in the thermodynamic limit $N\to \infty$. In the latter case, the integral is taken by the steepest descent
and its behaviour is analysed in more detail in the next section \ref{sec_exact}. To evaluate its behaviour for finite $N$ we 
will make use of the partition function zeros analysis, as outlined in sections  \ref{sec_method}, \ref{sec_zeros}.

\section{Phase diagram of the Blume-Capel model on the complete graph}\label{sec_exact}

\begin{figure}[h!]
\centerline{\includegraphics[angle=0, width=0.2\paperwidth]{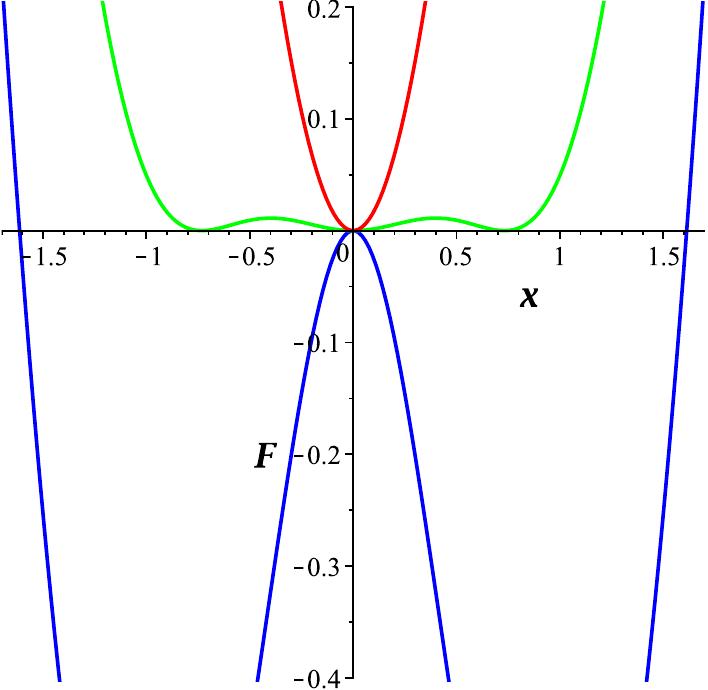}\hspace*{1cm}
\includegraphics[angle=0, width=0.2\paperwidth]{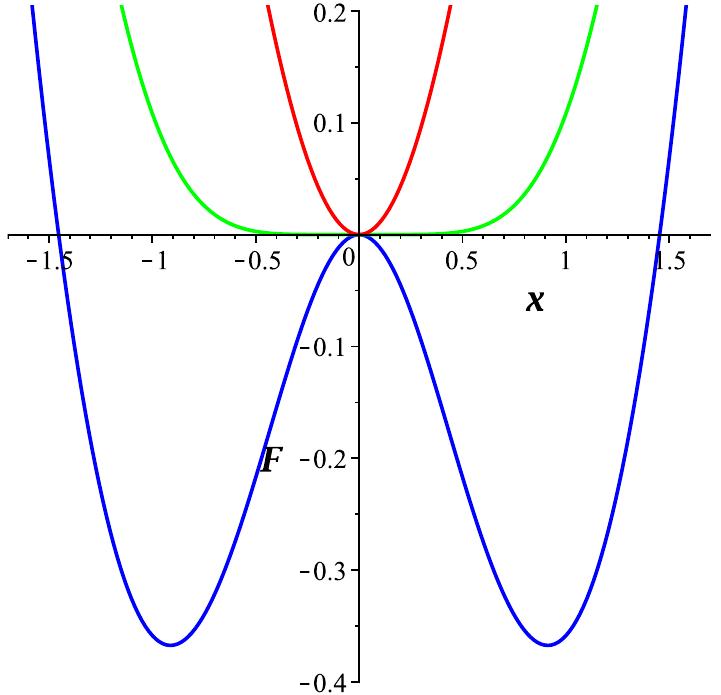}\hspace*{1cm}
\includegraphics[angle=0, width=0.2\paperwidth]{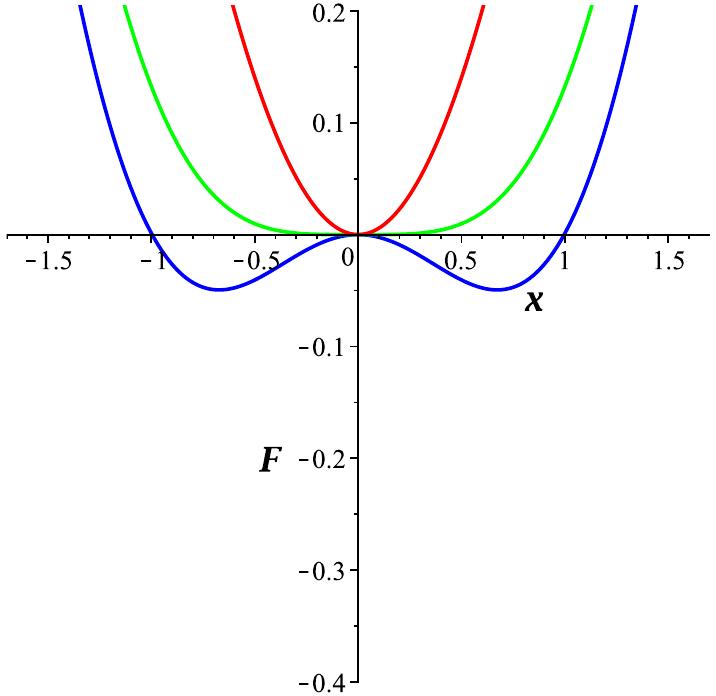}}
\centerline{{\bf (a)}  \hspace{10em} {\bf (b)} \hspace{10em} {\bf (c)}}
 \vspace*{1pt}
\caption{Behaviour of function $F(T,\Delta,x)$ (\ref{D2}) vs $x$ at {\bf (a)} $T=1/4$,
{\bf (b)} $T=T_t=1/3$, {\bf (c)} $T=1/2$ and different values of $\Delta$. In all figures the blue and the red curves correspond to
$\Delta=0.1$ and $\Delta=1$. The green curve corresponds to the first-order phase transition at $T=1/4$,
$\Delta=0.4762658366$ ({\bf a}), tricritical at $T=T_t=1/3$, $\Delta=\Delta_t=(2 \ln 2)/3=0.4620981204$ 
({\bf b}) and  the second order at $T=1/2$, $\Delta=0.3465735903$ ({\bf c})  transitions. The reference 
point for $F$ is  shifted by $\ln(1+2e^{-\Delta/T})$.
} \label{fig2hol}
\end{figure}

Let us discuss in more detail the phase diagram of the Blume-Capel model on the complete graph in the asymptotic limit $N\to \infty$. Consider the integral representation for the partition function, Eq. (\ref{II.4}):
\begin{equation}\label{D1}
Z_N(T,\Delta,H)= \int_{-\infty}^{+\infty} \exp\Big (-NF(T,\Delta,H,x)\Big)dx \, ,
  \end{equation}
where the expression in the exponential function  at $H=0$ reads:
\begin{equation}\label{D2}
F(T,\Delta,x) = \frac{x^2}{2T}-  \ln(1+2e^{-\Delta/T}
\cosh[x/T]) \, .
  \end{equation}
In the thermodynamic limit $N\to \infty$ the main contribution to the integral (\ref{D1}) is from the global minimum of $F(T,\Delta,x)$ as a
  function of $x$. In turn, the position of the minimum is governed by the pair of variables $(T,\Delta$). Depending on their values, one observes different types of behaviours as illustrated by Figs. \ref{fig2hol}. For low temperatures  
$T<1/3$ function $F$ has two symmetric  minima for low $\Delta$
(blue curve in Fig. \ref{fig2hol}{\bf a}). With an increase of $\Delta$  the function  
attains a specific three-minima shape and all minima get the same
value at a certain value of $\Delta$,  as shown by the green curve in Fig.
\ref{fig2hol}{\bf a}. Whereas only one minimum survives at $x=0$ for high $\Delta$
 (red curve in Fig. \ref{fig2hol}{\bf a}). This scenario differs for the intermediate
 region of temperatures $1/3<T\leq 2/3$ as illustrated in Fig. \ref{fig2hol}{\bf c}. 
 There, the two-minima shape of $F(x)$ at low $\Delta$ (blue curve)  transforms into a 
 single-minimum one with an increase of $\Delta$ (red curve) bypassing a three-minima behaviour. 
 The change in two regimes occurs at a certain value of $\Delta$ as
 shown by the green curve in Fig. \ref{fig2hol}{\bf b}. Finally, for $T>2/3$ only
 a single minimum is observed for $F(x)$ at any value of $\Delta$. Obviously, the scenarios
 observed at  $T<1/3$ and at $1/3 <T \leq 2/3$ describe the first and the second-order
 phase transitions, correspondingly, whereas a specific case when these two scenarios merge
 at $T=T_t=1/3$ describes at $\Delta=\Delta_t=(2 \ln 2)/3$ a tricritical point, as illustrated by the
 green curve in Fig. \ref{fig2hol}{\bf b}.

\begin{figure}[h!]
\centerline{\includegraphics[angle=0, width=8cm]{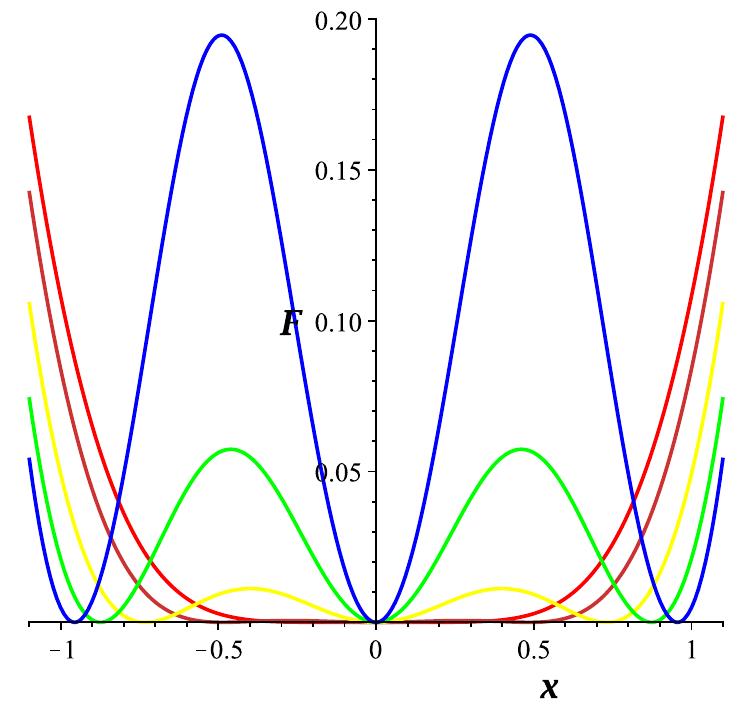}  \includegraphics[angle=0, width=8cm]{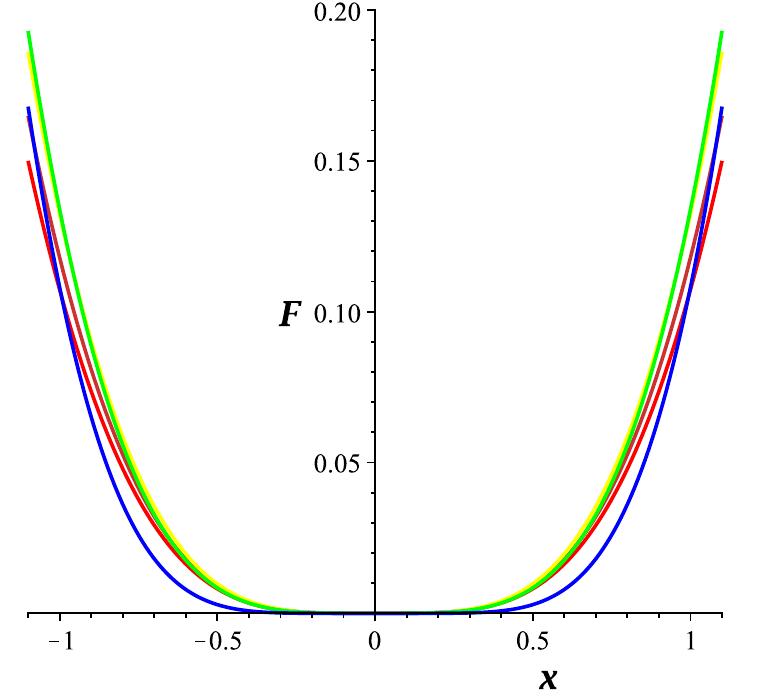} }
\centerline{{\bf (a)} \hspace{18em} {\bf (b)}  }
 \vspace*{1pt}
\caption{{\bf (a)} Behaviour of the function $F(T,\Delta,x)$ (\ref{D2}) vs $x$ at the tricritical point 
($T,\Delta$)= (1/3,(2$\ln(2)$)/3) (red curve) and at several points 
along the first-order phase transition line. ($T,\Delta$) = (0.3, 0.4667635075), (0.25, 0.4762658366), (0.2, 0.4867012711), (0.15, 0.4951587768),
orange, yellow, green, blue lines, correspondingly. 
{\bf (b)} Behaviour of  $F(T,\Delta,x)$ (\ref{D2}) as a function of $x$ 
at the tricritical point 
($T_t,\Delta_t$)= (1/3,(2$\ln(2)$)/3) (blue curve)
and at several points along the second-order phase
transition line: ($T,\Delta$) = (2/3,0), (0.6,0.1726092438), (0.5,0.3465735903), (0.4,0.4394449155), 
red, orange, yellow, green lines, correspondingly. 
The reference point for $F$ is shifted by 
$\ln(1+2e^{- \Delta/T})$.  
} \label{fig3hol}
\end{figure}

\begin{figure}[t!]
\centerline{\includegraphics[angle=0, width=8cm]{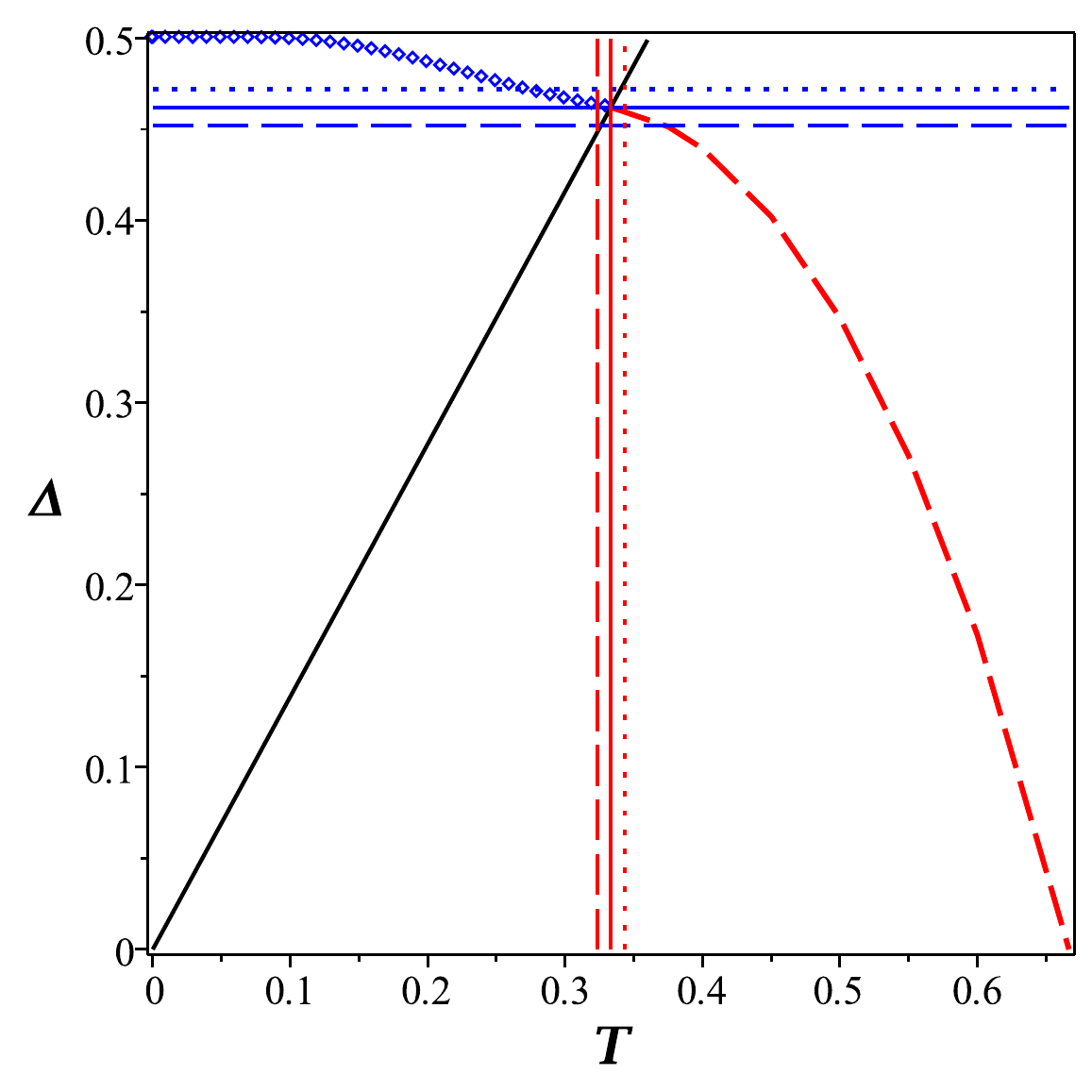} }
 \vspace*{1pt}
\caption{Phase diagram of the Blume-Capel model that follows from the analysis of function $F(T,\Delta,x)$ (\ref{D2}).
Blue diamonds: the first-order phase transition line, the dashed red line: the second-order phase transition line.
Both lines meet at the tricritical point ($T_t,\Delta_t$) = (1/3,(2$\ln(2)$)/3). The horizontal and vertical lines illustrate 
the approach to the phase transition we are going to analyse. 
The black line is given by Eq. (\ref{D6}), the derivative 
$\partial^4 F(T,\Delta,x) / \partial x^4$ vanishes along this line. Note that
$\partial^2 F(T,\Delta,x) / \partial x^2$ vanishes along the second-order phase transition line (the rightmost red dashed part of the phase diagram).} \label{fig6hol}
\end{figure}

Non-zero minima of $F(T,\Delta,x)$ that correspond to the first-order phase transition (see green curve in Fig.
\ref{fig2hol}{\bf a}) satisfy the system of equations:
\begin{eqnarray} \label{D3}
F(T,\Delta,x) - F(T,\Delta,0) = 0 \, , \\ \nonumber
 \partial F(T,\Delta,x) / \partial x = 0\, .   
\end{eqnarray}
For any fixed value of $T$ ($\Delta$) this system of equations defines the value of $\Delta$ ($T$)
at which the first-order phase transition occurs. The function $F(T,\Delta,x)$ at the first-order phase transition
for several pairs $T,\Delta$ is shown in Fig.  \ref{fig3hol}. One observes that the transition
becomes sharper with decreasing of $T$. Solving Eqs. (\ref{D3}) one gets a line of first-order phase 
transitions $\Delta(T)$ that terminates at the tricritical point ($T_t,\Delta_t$). This line is shown in Fig.
\ref{fig6hol} by blue diamonds (black solid curve in Fig. \ref{fig:phase_full}, correspondingly), and the
coordinates $(T,\Delta)$ of several points along the first-order 
transition line are given in Table \ref{tab1hol}.

\begin{table}[h!]
\caption{Coordinates $(T,\Delta)$ of several points along the first-order transition line, see  Figs.
\ref{fig:phase_full},
\ref{fig6hol}.
\label{tab1hol}}
\begin{center}
\footnotesize
    \begin{tabular}{|c|c|c|c|c|c|c|}
     \hline 
       $T$ & 0 & 0.05 & 0.1 & 0.2 & 0.3 & 0.32 \\  
       \hline 
            $\Delta$ & 1/2 & 0.4999977309 & 0.4993435185 & 0.4867012711 &  0.4667635075 & 0.4624864590 \\
      \hline
    \end{tabular}
\end{center}
\end{table}

In the second-order phase transition region at $T>T_t$, the minimum of $F(T,\Delta,x)$ at the transition
point (see green curve in Fig. \ref{fig2hol}{\bf c}) satisfies
\begin{equation} \label{D4}
 \partial^2 F(T,\Delta,x) / \partial x^2 = 0\, .   
\end{equation}
This equation defines the $\Delta(T)$ second-order phase transition line, shown by the black dashed curve
in Fig. \ref{fig:phase_full} and by the red dashed curve
in Fig. \ref{fig6hol}. With an explicit expression for the function $F(T,\Delta,x)$ to hand and taking into account
that the derivative in Eq. (\ref{D4}) is taken at the extremum $x=0$ one arrives at the following equation
for the second-order phase transition line:
\begin{equation} \label{D5}
 T\Big (e^{\Delta/T} +2 \Big ) = 2\, .   
\end{equation}
Coordinates $(T,\Delta)$ of several points along this line are given  in Table \ref{tab2hol}. The line terminates
at the critical point of the $S=1$ Ising model: $(T,\Delta)=(2/3,0)$. The second-order and the first-order
phase transition lines meet at the tricritical point
($T_t,\Delta_t$) = (1/3,(2$\ln(2)$)/3). The behavior of $F(T,\Delta,x)$ as a function of $x$ at the tricritical 
point ($T_t,\Delta_t$) is shown in  Fig. \ref{fig3hol}{\bf a} by red and in Fig.  \ref{fig3hol}{\bf b} by blue lines.

\begin{table}[h!]
\caption{Coordinates $(T,\Delta)$ of several points
along the second-order transition lines, dashed black and red curves in Figs.
\ref{fig:phase_full}, \ref{fig6hol}, correspondingly.
\label{tab2hol}}
\begin{center}
\footnotesize
    \begin{tabular}{|c|c|c|c|c|c|c|}
     \hline 
       $T$ & 0.34 & 0.4 & 0.5 & 0.6 & 0.66 & 2/3 \\  
       \hline 
            $\Delta$ & 0.4611900752 & 0.4394449155 & 0.3465735903 & 0.1726092438  & 0.01970295557 & 0 \\
      \hline
    \end{tabular}
\end{center}
\end{table}

Note that at the tricritical point both the second and the fourth derivative of $F(T,\Delta,x)$ in $x$ vanish.
Therefore this point can be also defined as a crossing point of two lines, one defined by condition (\ref{D4}),
and given by function (\ref{D5}), the other one defined by the condition 
\begin{equation} \label{D6}
 \partial^4 F(T,\Delta,x) / \partial x^4 = 0\, .   
\end{equation}
Solving Eq. (\ref{D6}) one arrives at
\begin{equation} \label{D7}
 \Delta=2 T \ln 2 \, .   
\end{equation}
This line in shown in Fig. \ref{fig6hol} in black. In the following, we will be interested in
studying the partition function zeros for  finite values of the number of sites $N$ when the phase transition is approached at $T=T_t$ or
$\Delta=\Delta_t$ (solid red and blue lines in Fig. \ref{fig6hol}) as well as in the vicinity 
of the tricritical point (moving along dashed and dotted straight lines of Fig. \ref{fig6hol}) 
and of the second-order phase transition line, Eq. (\ref{D5}). To study the effective critical and
tricritical behaviours of finite-size systems, we will rely on the analysis of partition function zeros,
as further outlined below.

\section{Theoretical aspects of the partition function zeros analysis}\label{sec_method}

The partition function of a finite system, as a function of real external parameters, cannot reach zero value since no singularity of the free energy can be observed except in the thermodynamic limit. However, that is not the case if physical parameters (like $T$, $H$, or $\Delta$) are considered complex. 
The partition function formalism for the analysis of critical behaviour is useful especially because it allows to investigate a wide range of systems either analytically \cite{Pearson82,itzykson_distribution_1983, glasser_complex_1986,glasser_complex-temperature-plane_1987,kenna_scaling_2006,krasnytska_violation_2015,krasnytska_partition_2016,e26030221} or numerically  \cite{janke_strength_2001,RuizLorenzo24,moueddene_critical_2024,PhysRevE.110.064144}.  It was firstly proposed by Lee and Yang in 1952  for the lattice gas model, and later used as a method to consider the magnetic field as a complex variable and investigate the partition function representation as a polynomial in a related parameter \cite{lee_statistical_1952,yang_statistical_1952}.
Then the method was extended to the analysis of the zeros  in the complex temperature plane by Fisher  \cite{Fisher68}. Universal characteristics can be obtained from the coordinates of the zeros in the corresponding complex plane. This characteristics are connected by a specified universal relation with critical exponents and amplitudes.  Fisher zeros are accumulated in the vicinity of the (real valued) critical point $T_c$ along a line and tend to pinch the positive real axis with the angle $\varphi$. There exist a  relation that connects the impact angle $\varphi$ with the critical
exponent of the specific heat $\alpha$  and with the specific heat universal amplitude ratio $A_-/A_+$, where $A_+$, $A_-$ are  scaling
amplitudes at $T>T_c$, $T<T_c$ \cite{Pearson82, kenna_scaling_2006,
itzykson_distribution_1983}: 
 \begin{equation}\label{I.3}
\tan[{(2-\alpha)\varphi}]=\frac{\cos(\pi \alpha) -A_-/A_+}{\sin(\pi
\alpha)}.
\end{equation}  

The scaling relation of the zeros coordinates follows from general arguments and for Lee-Yang and Fisher zeros read \cite{itzykson_distribution_1983, kenna_scaling_2006,krasnytska_partition_2016,BERCHE2012115}:
\begin{equation}\label{I.10}
	H_j(N,T=T_c)\sim  \left(  \frac{j}{N}
	\right)^{g_h}\,  , \hspace{1cm}
	T_j(N,H=0)\sim  \left(  \frac{j}{N}  \right)^{g_t}, \hspace{1cm}
    g_h=\frac{\beta\delta}{2-\alpha}\,  , \hspace{1cm}
	 g_t=\frac{1}{2-\alpha},
\end{equation}
where label $j$ enumerates the order of zeros. Exponents $g_h, g_t$, called gap exponents since they show the gap between the partition function zero and real axis, are usually denoted with the letter $\Delta$, are changed here to not be confused with the crystal field.

Since the critical and tricritical singularities of the Blume-Capel model on the complete
graph in the asymptotic limit $N\to \infty$ are governed by the mean-field exponents, one can derive
the values for the angles and exponents that describe zeros behaviour on their basis. These values
are summarized in Table \ref{tab_expectedvalues}.

\begin{table}[h!]
    \centering
    \begin{tabular}{|c|c|c|}

       \hline 
       Known MFA values \cite{butera_blumecapel_2018,rocha-neto_thermodynamical_2023,10.21468/SciPostPhysLectNotes.60} & Tricritical point ($T_{t}$) & Critical point ($T_c$) \\ \hline
        $\alpha$ & 1/2   & 0\\
        $\beta$  & 1/4   & 1/2\\
        $\delta$ & 5 & 3\\
        $\gamma$ & 1   & 1\\
        $A_+/A_-$ & 0 & 0\\ \hline
         Expected  results & & \\ \hline
          $\varphi$, see Eq.(\ref{I.3}) & $\pi/3$ & $\pi/4$\\ 
           $g_h$ from Eq. (\ref{I.10}) & 5/6 & 3/4\\
          $g_t$ from Eq. (\ref{I.10}) & 2/3 & 1/2\\
       \hline
    \end{tabular}
    \caption{Critical exponents and expected scaling for zeros' coordinates and angles of the Blume-Capel model at tricritical and critical points.}
    \label{tab_expectedvalues}
\end{table}

For the Blume-Capel model, in addition to the Lee-Yang and Fisher zeros, one can make use of the analysis of
the partition function zeros in a complex crystal field plane at zero magnetic field \cite{biskup_general_2000, moueddene_critical_2024}. Given a point on 
the critical line (see Fig.  \ref{fig:phase_full}), we substitute $\Delta={\rm Re}\,\Delta+i\,{\rm Im}\,\Delta$. 
It is argued that the behaviour of these Crystal Field zeros is to be expected in line with that of the Fisher zeros, 
with the only exception of the fact that the scaling field connected to the crystal field is well-defined 
at the tricritical point and behaves poorly as it reaches the ``Ising'' critical point $\Delta=0$ \cite{moueddene_critical_2024}.  
This issue will be investigated in more detail in this paper later. 

In what follows below, we investigate the behaviour of the partition function zeros in different regions of complex parameters. 
We will consider the crossover between different regimes and scaling with system size $N$. In particular, we will investigate 
the partition function zeros (Fisher, Crystal Field and Lee-Yang zeros)  in the vicinity of the critical point, near the critical 
line and near the tricritical point. 

\section{Zeros analysis for the exact partition function}\label{sec_zeros}

For further investigation, we use the exact expression for the partition function (\ref{II.4}) in different regions of the parameters $T,\Delta, H$ and find zeros' characteristics numerically for different system sizes.

\subsection{Fisher zeros }\label{results_exact}

 Substituting $T ={\rm Re}\, T+i\, {\rm Im}\, T $ into the partition function at zero magnetic field, we get:
\begin{equation} \label{main}
Z_N (T_R+iT_I,\Delta,H=0)= \int_{-\infty}^{+\infty} \exp\Big \{ -N \big [\frac{(T_R+iT_I)x^2}{2}- \ln(1+2e^{-(\frac{1}{2N}+ \Delta)/(T_R+iT_I)}
\cosh x)\big ]\Big \} dx \, .
  \end{equation}

We are interested in the simultaneous solutions of the equations 
\begin{equation}
    {\rm Re}\,  Z_N (T_R+iT_I,\Delta,H=0) = 0 \quad \hbox{and}\quad {\rm Im}\,  Z_N (T_R+iT_I,\Delta,H=0) = 0.
\end{equation} 
We calculate the zeros of a given partition function (or intersections of the curves described by the above equations) as we move from the critical point along the critical line to the tricritical point.
\subsubsection{Near the ``Ising'' critical point: $\Delta_c=0$, $T_c=2/3$, $H_c=0$} \label{results_exact_fisher_criticalpoint}
\begin{figure}[h!]
    \centering
    \includegraphics[width=0.7\linewidth]{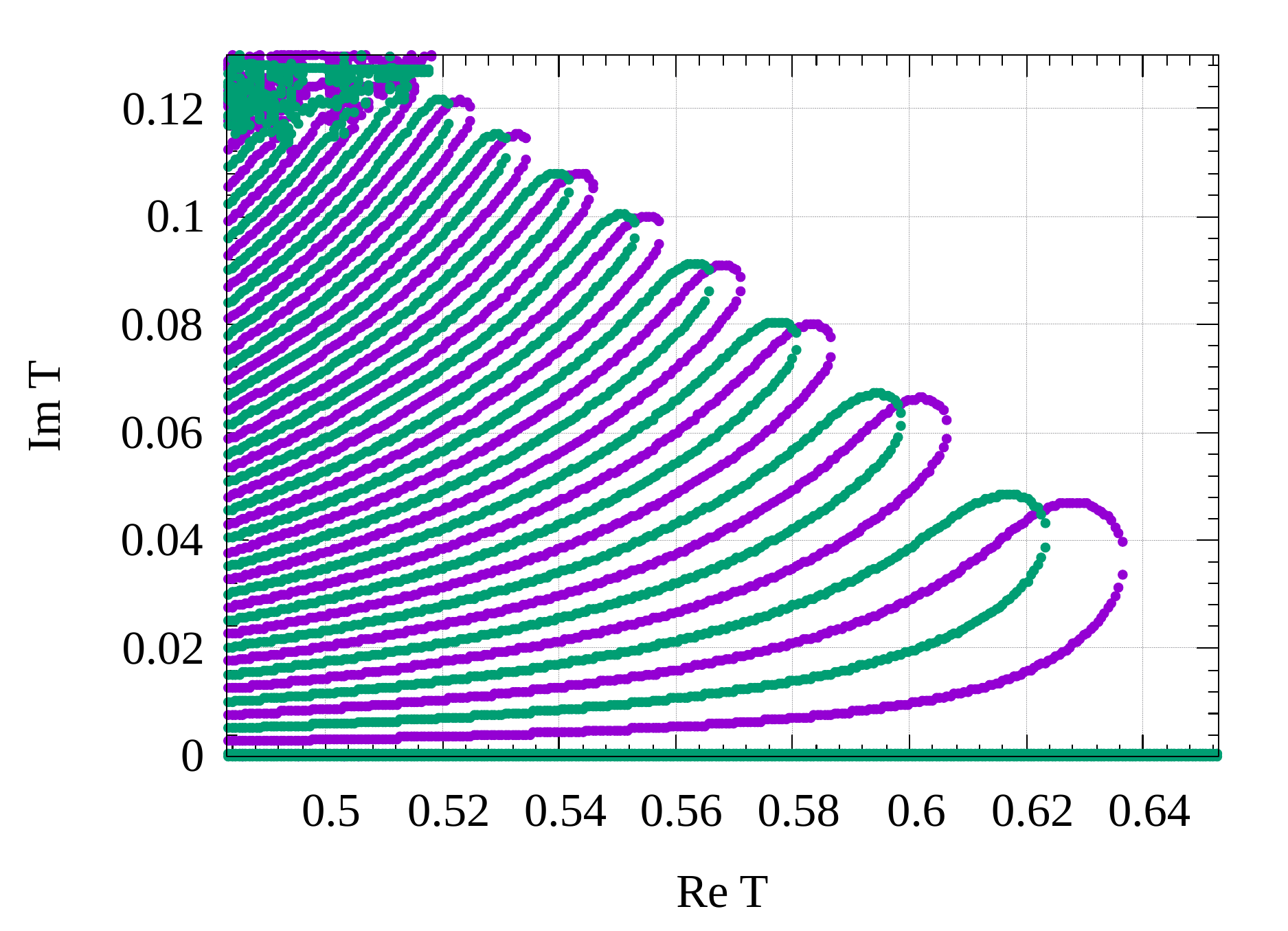}
    \caption{Solutions of the equations ${\rm Re}\,  Z_N (T_R+iT_I,\Delta=0,H=0) = 0$ (purple curves) and ${\rm Im}\,  Z_N (T_R+iT_I,\Delta=0,H=0) = 0$ 
    (green curves) in the complex temperature plane for $N=500$ in the vicinity of  $T_c$ ("Ising" point). Fisher zeros are obtained by seeking the intersections of 
    green and purple curves. Same colours are kept throughout the paper.}
    \label{fig:Fish_cri_zeros}
\end{figure}

\begin{figure}[h!]
    \centering
    \includegraphics[width=0.8\linewidth]{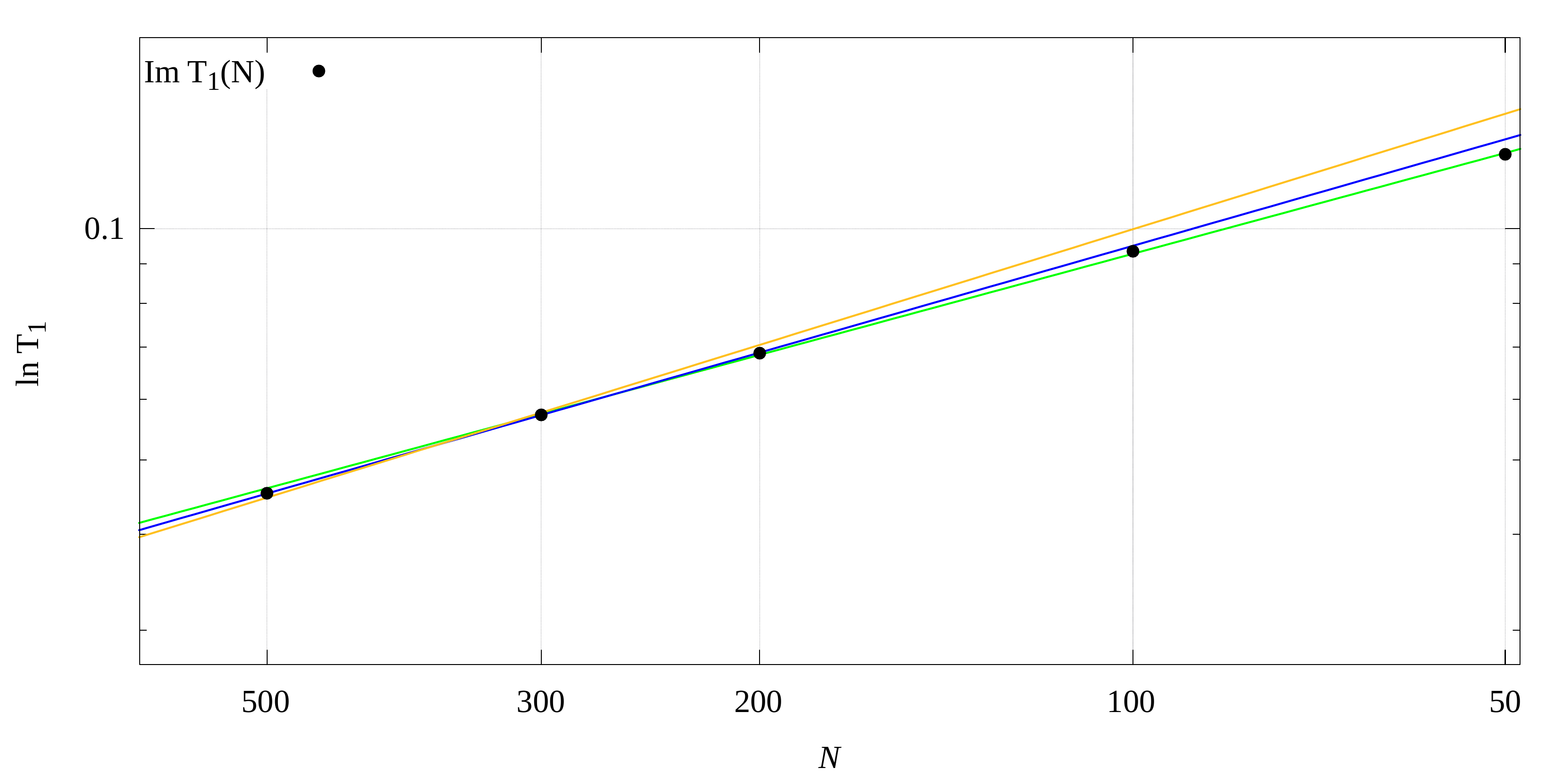}
    \caption{Coordinate of the first Fisher zero $T_1$ as a function of system size $N$. The expected scaling (yellow line) is $T_1\sim(1/N)^{1/2}$. For the graph sizes up to $N=500$ we obtain $T_1\sim(1/N)^{0.44}$ or $T_1\sim(1/N)^{0.46}$ if we only take into account three largest system sizes, which shows that the scaling approaches the theoretical value, even though slowly.}
    \label{fig:Fish_cri_fss}
\end{figure}

At this point on the symmetric $T-\Delta$ plane the Fisher zeros are well defined, see Fig.~\ref{fig:Fish_cri_zeros}. The intersections of lines where real and imaginary parts of the partition function vanish can be clearly seen even at low resolution. Yet the finite-size scaling (Fig.~\ref{fig:Fish_cri_fss}),  is still not at the expected values. We expect, as the system size $N$ rises, the first Fisher zero to scale as $T_1\sim N^{1/2}$, yet we only observe $T_1 \sim N^{0.44}$ for all system sizes, and $T_1\sim N^{0.46}$ when only the largest graph sizes were taken into account. 

\subsubsection{Near the critical line:  $\Delta< 2/3 \ln 2$,  $T<2/3$, $H=0$} \label{results_exact_fisher_criticalLine}
\begin{figure}[h]
    \centering
    \includegraphics[width=0.45\linewidth]{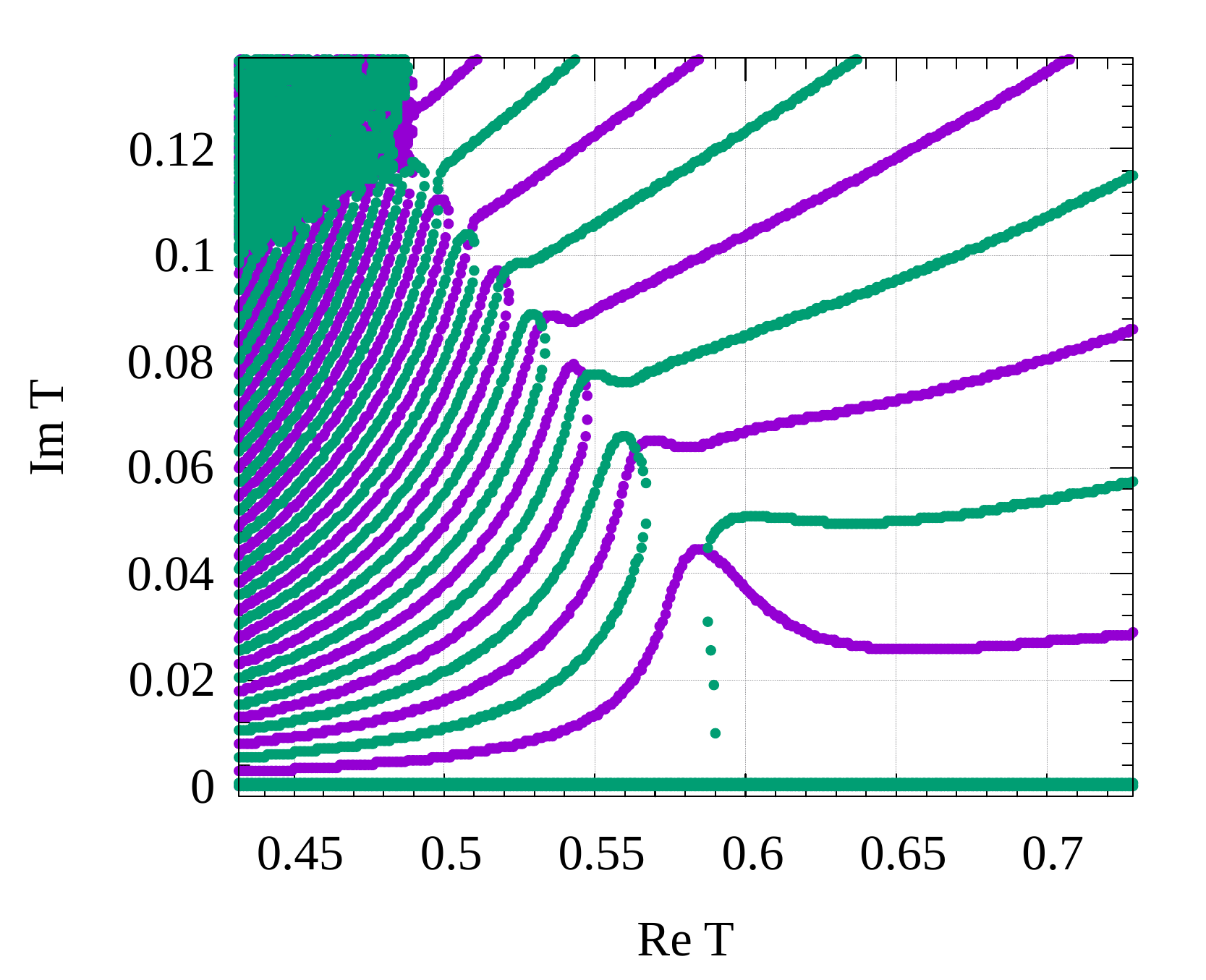} \includegraphics[width=0.45\linewidth]{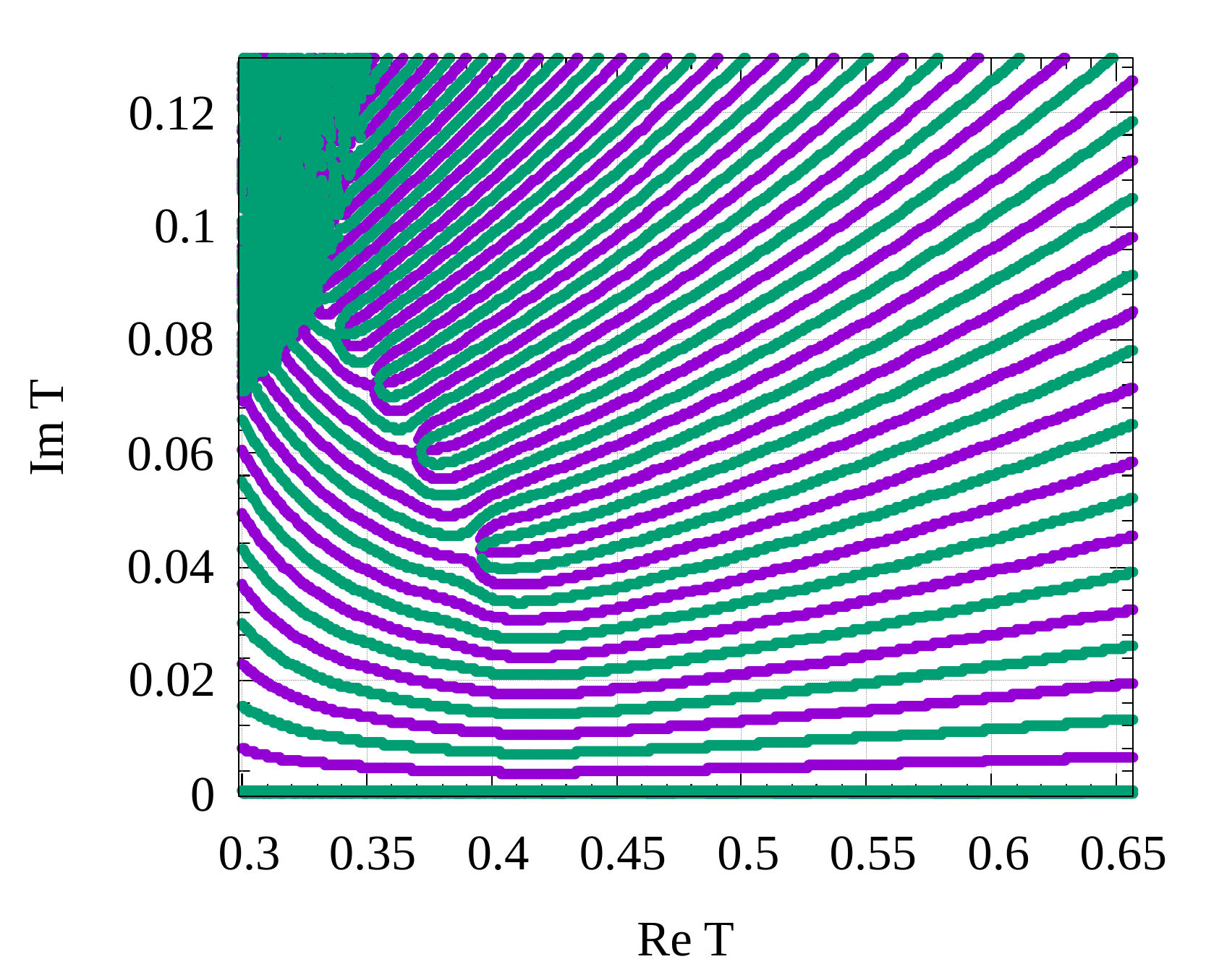}
    \caption{Fisher zeros for $N=500$ along the critical line at $\Delta=0.1$ (left) and near $T_t$ (right). As we approach the tricritical point, the lines  ${\rm Re}\,  Z_N (T_R+iT_I,\Delta,H=0) = 0$ and ${\rm Im}\,  Z_N (T_R+iT_I,\Delta,H=0) = 0$ become more dense, and the amplitudes of the functions are more substantial, meaning the impossibility to distinguish a zero of the partition function.}
    \label{fig:Fish_near_tri_zeros}
\end{figure}

As we pick different points along the critical line  as the points approach the tricritical point, we see the lines of the zeros first unfold (Fig.~\ref{fig:Fish_near_tri_zeros}, left) and then completely flip its direction (Fig.~\ref{fig:Fish_near_tri_zeros}, right) in comparison to their orientation at the "Ising" critical point. Above a certain level, the amplitudes of the functions ${\rm Re}\, Z$ and ${\rm Im}\, Z$ drastically increase, which, in combination with the zero lines becoming more dense, leads to the cloud of almost random results (seen in both plots in the top left corner) that become much stronger as the tricritical point is approached, and a lesser total number of distinct Fisher zeros can be  defined. The results obtained from the zeros along the critical line are discussed later in this Section.

 \subsubsection{Near the tricritical point: $\Delta_t=2/3\ln 2$, $T=T_t=1/3$, $H_{t}=0$}\label{results_exact_fisher_Tricritical}
\begin{figure}[h]
    \centering
    \includegraphics[width=0.7\linewidth]{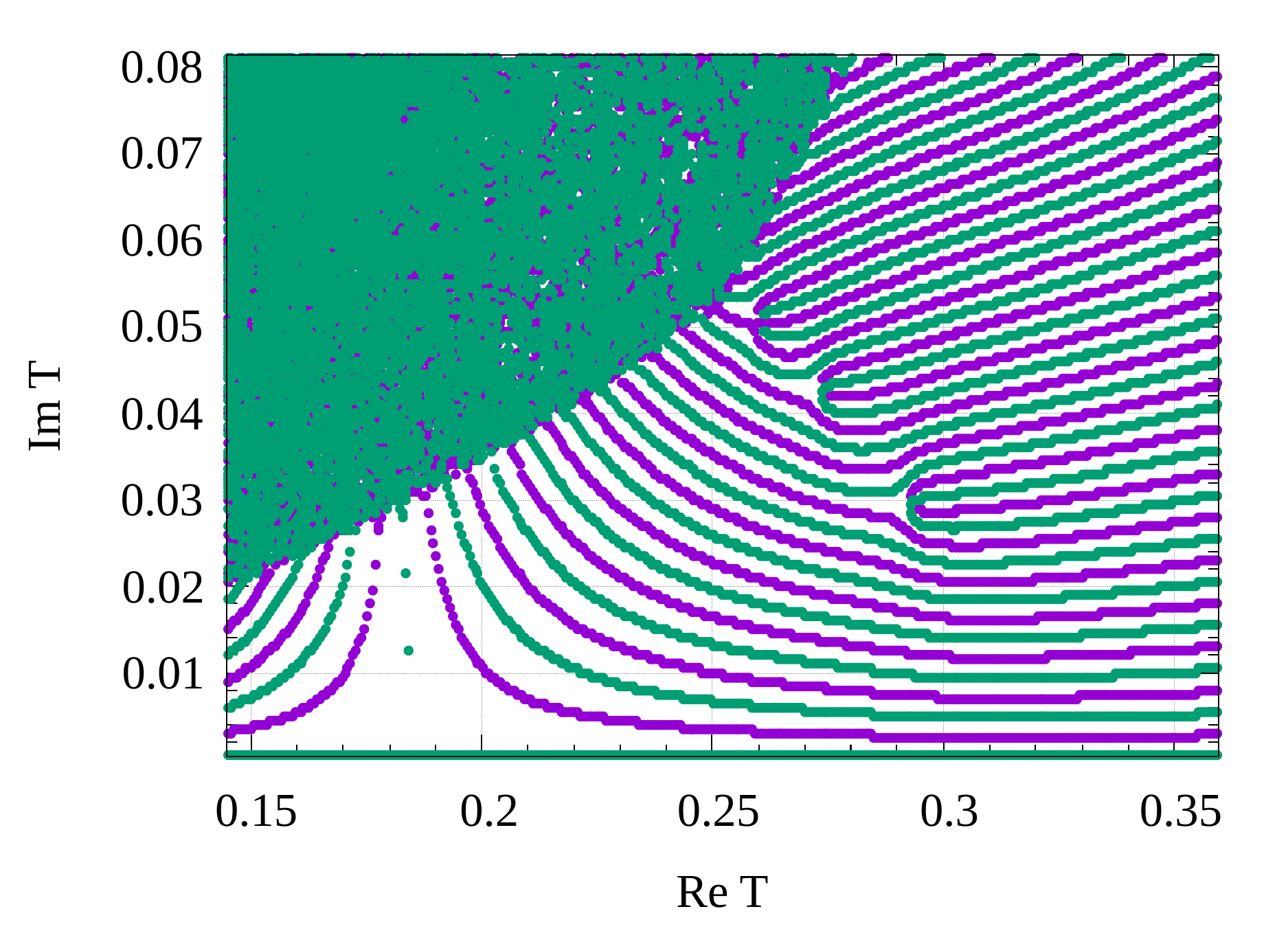}
    \caption{Fisher zeros for $N=500$ at the $T_t = 1/3$. Only a few first zeros can be seen, while the rest hides behind the limits of our computing capabilities.}
    \label{fig:Fish_tri_zeros}
\end{figure}

\begin{figure}[h]
    \centering
    \includegraphics[width=0.7\linewidth]{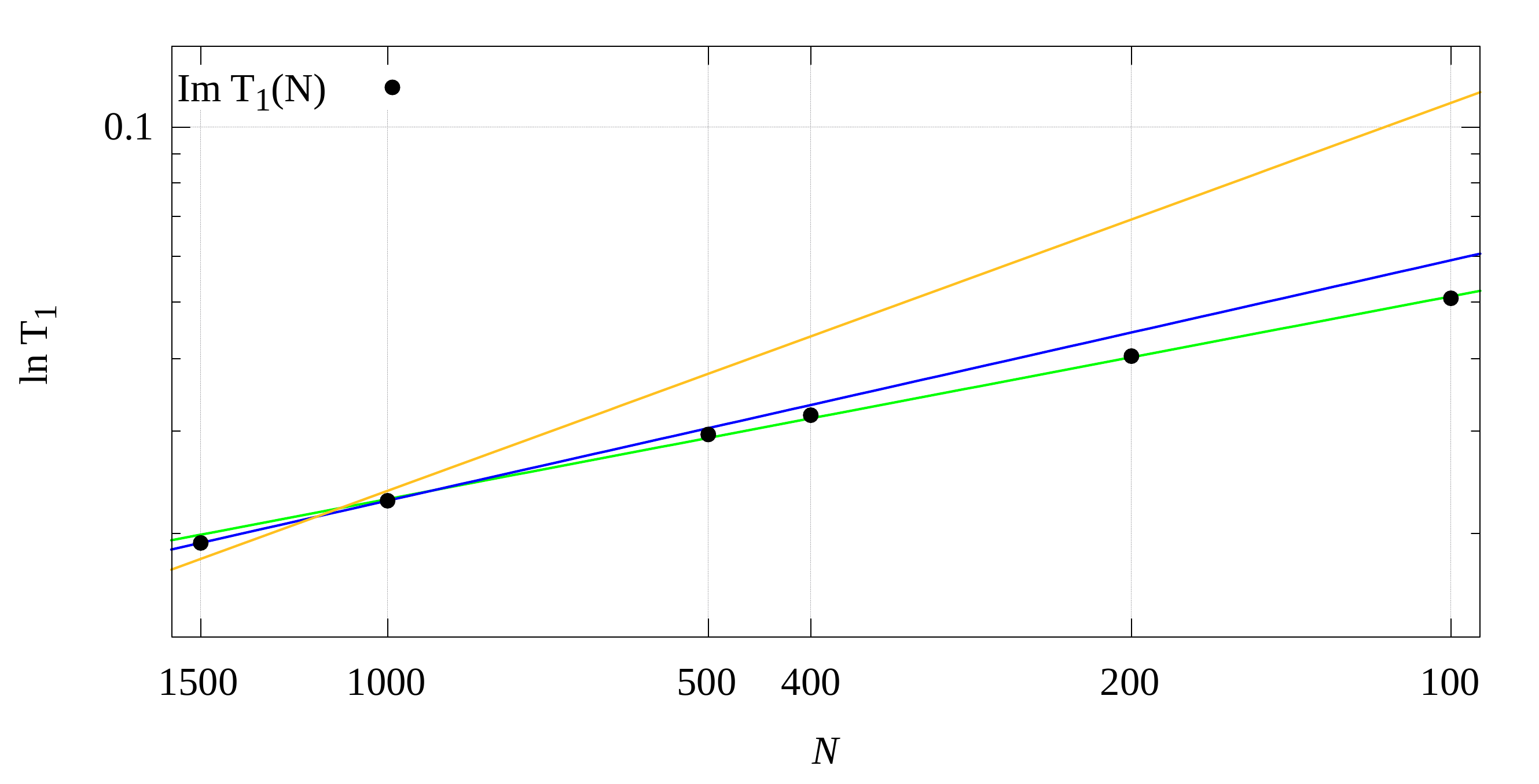}
    \caption{Coordinate of the first Fisher zero $T_1$ as a function of system size $N$    
    near $T_t=1/3$ at $\Delta_t = 2/3 \ln2$. The mean-field expectation is $T_1\sim(1/N)^{2/3}$, while for the graph sizes up to $N=1500$ we obtain $T_1\sim(1/N)^{0.35}$ (green line) or $T_1\sim(1/N)^{0.41}$ if we only take into account three largest system sizes, which shows that the asymptotic 
    scaling approaches very slowly and the effective finite-size behaviour is of great importance at the tricriticality.}
    \label{fig:Fish_tri_FSS}
\end{figure}

At the tricritical point the zero lines in the complex temperature plane become so dense and have such large amplitudes, that just a few first Fisher zeros are visible (Fig.~\ref{fig:Fish_tri_zeros}),  and the FSS at our largest system sizes $T_1\sim N^{0.41}$ is very far from the expected $T_1\sim N^{0.66}$, see Fig.~\ref{fig:Fish_tri_FSS}.

\subsubsection{Impact angle analysis}

\begin{figure}[h]
    \centering
    \includegraphics[width=0.7\linewidth]{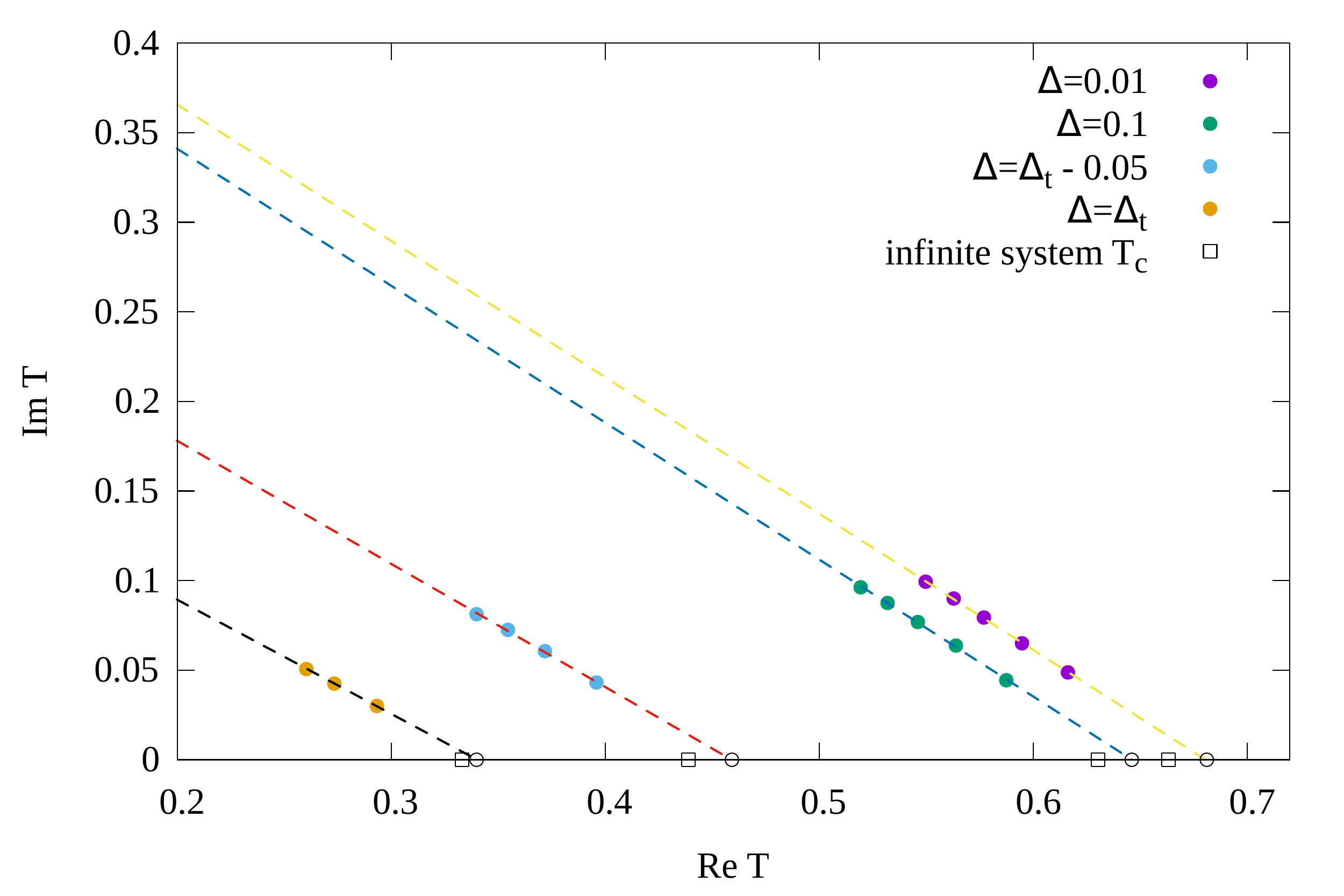}
    \caption{Fisher zeros for $N=200$ at different points along the critical line ($\Delta = 0.01, 0.1, \Delta_t-0.05$) and at the 
    tricritical point $\Delta_t=2/3 \ln2$. The fitted angle values read $\varphi = 37.27, 37.38, 34.48, 32.51$ correspondingly. Note that angles $\varphi$ should be equal ($\varphi_{MF}=45^{\circ}$) for the infinite 
    system at all points along the critical line and become $\varphi_{MF}=60^{\circ}$ at the tricritical point.}
    \label{fig:Fish_cri_line_angles}
\end{figure}

Fisher zeros align in a way that as they approach the real axis, they form an angle with its negative direction. The value of this angle is connected to scaling exponents, as well as to the ratio of amplitudes of the specific heat, see Eq.(\ref{I.3}). In the thermodynamic limit, we expect the point of intersection to be the critical temperature at that crystal field value. However, for a finite graph, the point of intersection would lie somewhere else, and it can be argued that it depicts an effective phase change point (pseudo-critical point). Thus, we fix several values of the crystal field, and study how the impact angle and the pseudo-critical point at finite size differ from their infinite system counterparts, see e.g. Fig.~\ref{fig:Fish_cri_line_angles} for such comparison for a complete graph of size $N=200$. The intersection point seems to  be positioned always at a higher temperature than the critical point of the infinite system. 

\begin{figure}[h]
    \centering
    \includegraphics[width=0.7\linewidth]{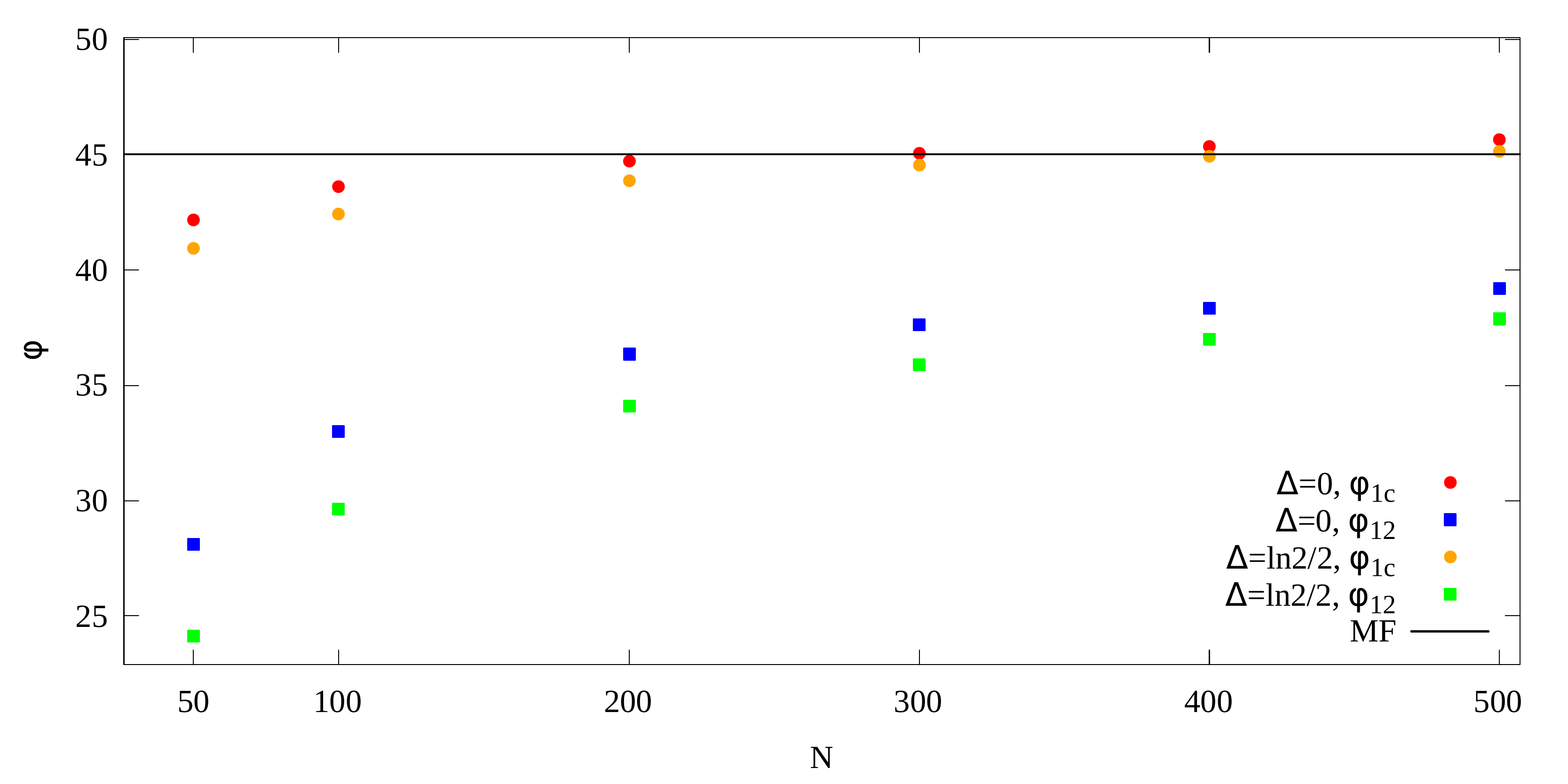}
    \caption{Angles $\varphi_{1c}$ and $\varphi_{12}$ between the Fisher zeros and real temperature axis for different system sizes at $\Delta=0$ and along the critical line at $\Delta=\ln2 / 2$. In the thermodynamic limit both angles should be $\varphi_{MF}=45^\circ$, and we can already see that for graphs up to the size $N=500$ angle $\varphi_{1c}$ reaches the asymptotics.}
    \label{fig:Fish_plot_angles}
\end{figure}

We also want to observe whether the impact angle approaches the scaling value. There are two ways to calculate this angle, one is looking for an angle
between the real axis and direction from the first zero to the critical point. This angle will be denoted as $\varphi_{1c}$. Another way is to define an
angle between the real axis and the direction from the second zero to the first zero, further denoted as $\varphi_{12}$. As we can see from
Fig.~\ref{fig:Fish_plot_angles}, the former angle approaches the asymptotic value faster, while the latter has not reached the expected angle value 
for the sizes that we were able to compute.

 \clearpage


\subsection{Crystal field zeros}\label{results_exact_crystal}
Let us now analyse the partition function zeros in the complex crystal field plane.
To this end we substitute into Eq. (\ref{II.4}) $\Delta = \Delta_r + i\Delta_i$ \cite{lawrie_theory_1984} resulting
in:
\begin{equation} \label{main_crystal_field}
Z_N (T,\Delta_r + i\Delta_i,H=0)= \int_{-\infty}^{+\infty} \exp\Big \{ -N \big [\frac{Tx^2}{2}- \ln(1+2e^{-(\frac{1}{2N}+ (\Delta_r + i\Delta_i))/T}
\cosh x)\big ] \Big \} dx \, .
  \end{equation}

The shape of the phase diagram near a tricritical point is intimately linked to the
critical exponents
governing the singular behaviour there. We assume that the equations of the
critical line and the first-order phase transition line near the tricritical point are, respectively:
\begin{eqnarray}
    \Delta_2 (T) = \Delta_t - a(T-T_t) + b_2 (T-T_t)^{\psi}+\dots,  \\
    \Delta_1 (T) = \Delta_t - a(T-T_t) + b_1 (- T+T_t)^{\psi_t}+\dots,
\end{eqnarray}
where the common coefficient $a$ ensures that both lines have the same slope at the tricritical point. The exponents $\psi$ and $\psi_t$ are shift exponents, describing the shape of the phase boundary near the tricritical point.

Another scaling field $g$ is introduced such that it is perpendicular to the tangent to the phase separation line at the tricritical point:
\begin{equation}
    g = \Delta - \Delta_t - a(T-T_t).
\end{equation}
This way, the expected scaling for the crystal field zeros coincides with the scaling for Fisher zeros, meaning that in the same way as the first Fisher zero, first crystal field zero would scale according to:

\begin{equation}
    \Delta_1\sim \left(\frac1{N}\right)^{g_t}, \quad g_t = \frac1{2-\alpha}.
\end{equation}

\begin{figure}[h!]
    \centering
    \includegraphics[width=0.5\linewidth]{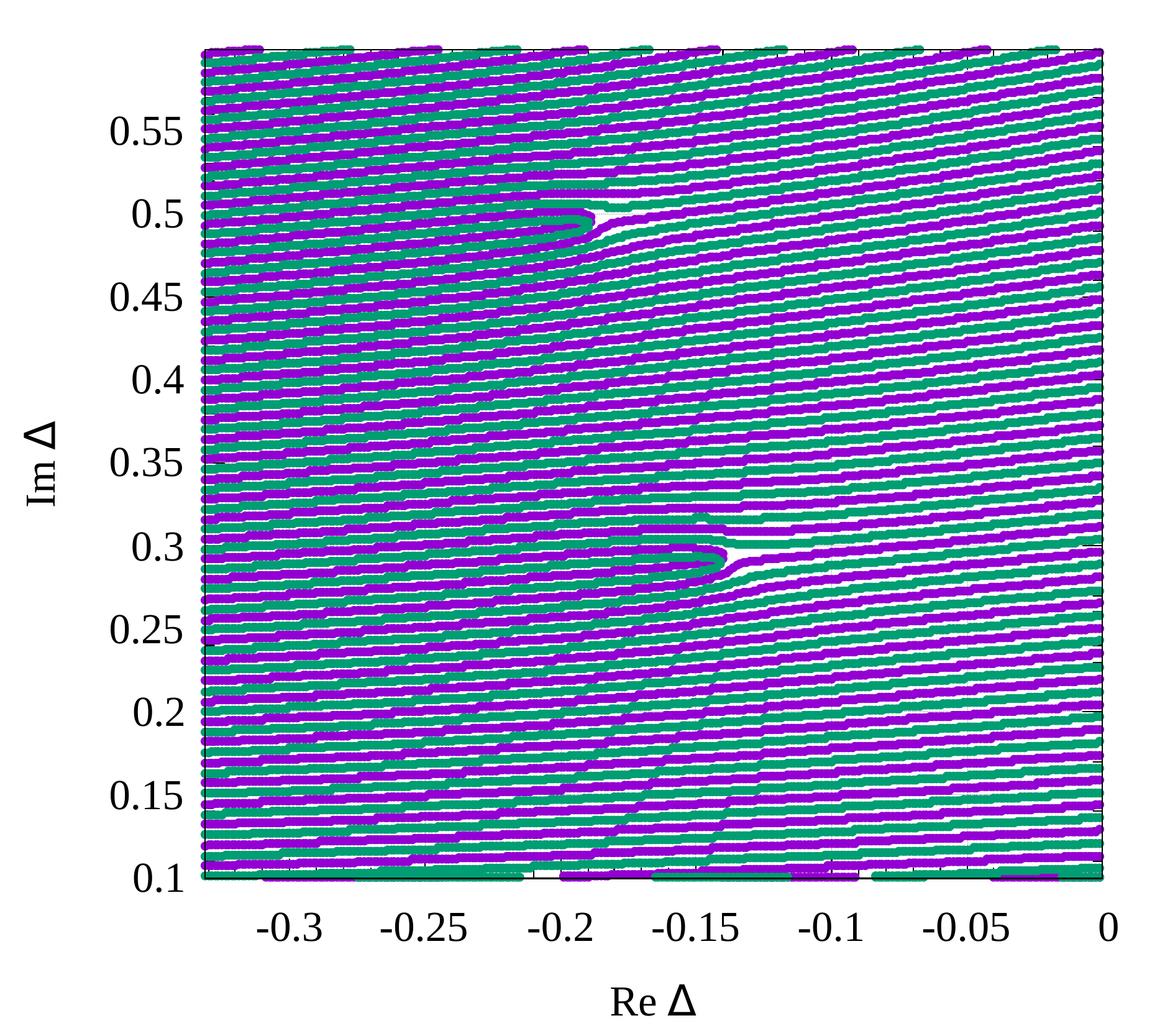}
    \caption{Solutions of equations ${\rm Re}\, Z=0$ and ${\rm Im}\, Z=0$ in complex crystal field plane for $N=200$ at $T_c=2/3$. Crystal field zeros are difficult to spot due to the high density of zero lines. This happens due to the fact that the critical line here is not perpendicular to the scaling field.}
    \label{fig:Cryst_cri}
\end{figure}

 \subsubsection{"Ising" critical point: $\Delta_c=0$, $T_c=2/3, H_c=0$} \label{results_exact_crystal_criticalpoint}
At this point on the symmetric $T-\Delta$ plane the crystal field zeros are not well defined, see Fig.~\ref{fig:Cryst_cri}. This occurs regularly \cite{moueddene_critical_2024,PhysRevE.110.064144} because the scaling field $g$ is directed perpendicularly to the tricritical point, and not to the critical point.


\subsubsection{Near the critical line:  $\Delta< 2/3 \ln 2$,  $T<2/3$, $H=0$} \label{results_exact_crystal_criticalLine}

\begin{figure}[h!]
    \centering
    \includegraphics[width=0.8\linewidth]{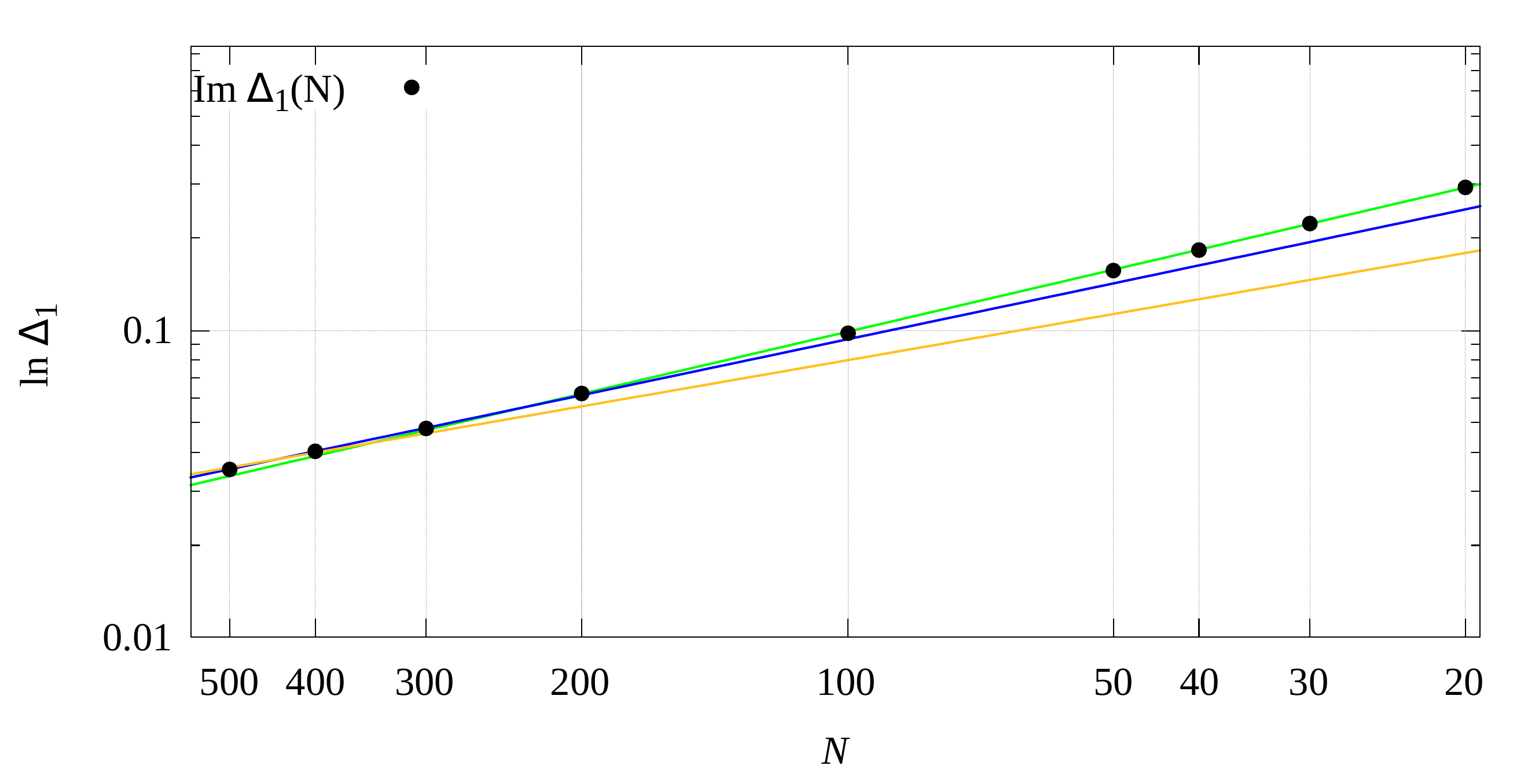}
    
    \includegraphics[width=0.8\linewidth]{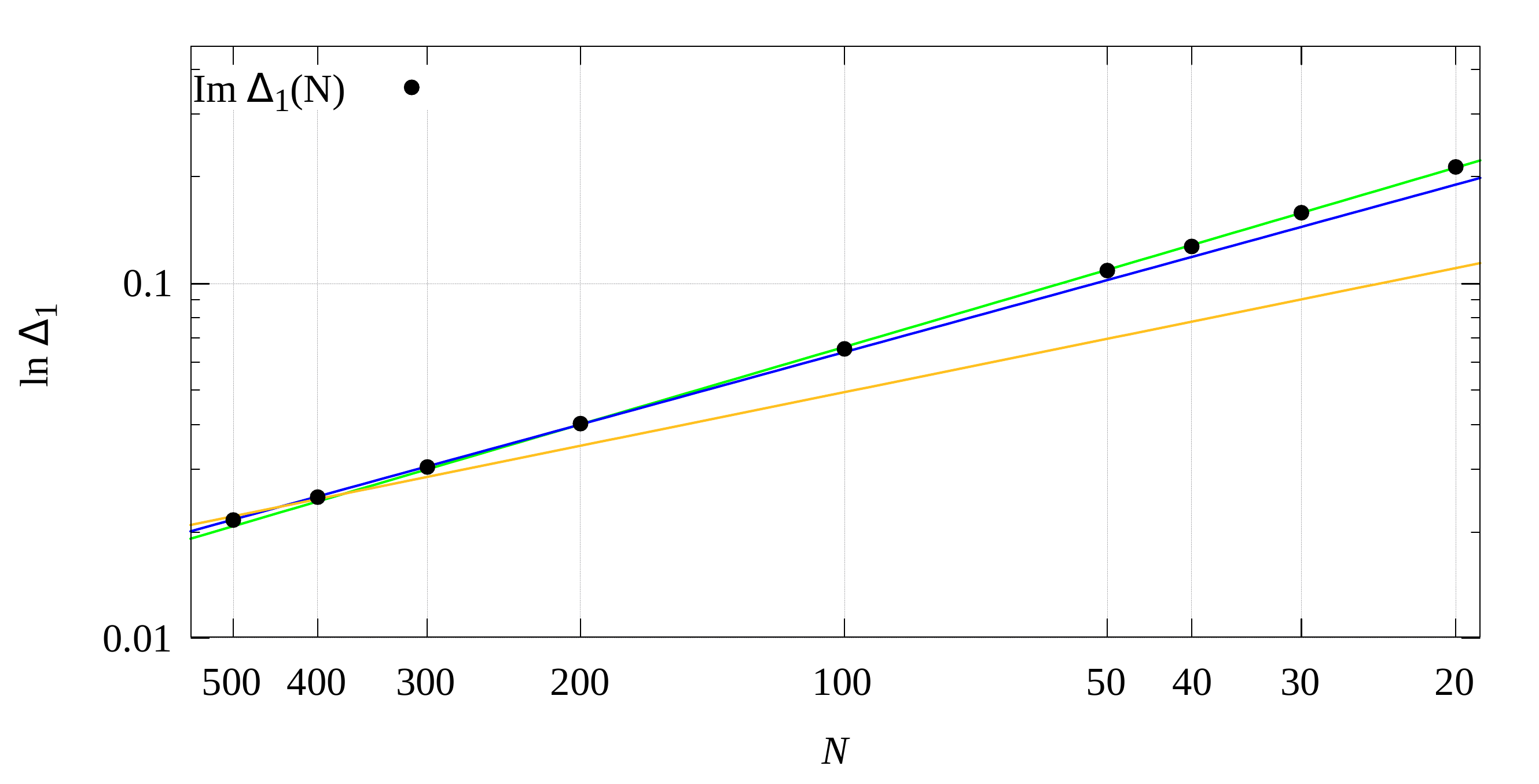}
    \caption{Coordinate of the first crystal field zero $\Delta_1$ as function of $N$ at $T = 0.4$ (top) and $T=0.35$ (bottom) for graph sizes up to $N=500$. 
    While the expected mean-field scaling would be $\Delta_1\sim N^{1/2}$, we observe $\Delta_1\sim N^{0.62}$ and $\Delta_1\sim N^{0.69}$ respectively, 
    much closer to the tricritical behaviour $\Delta_1\sim N^{2/3}$.}
    \label{fig:Cryst_cri_line_fss}
\end{figure}

Along the critical line, as we approach the tricritical point, the crystal field zeros and their finite-size scaling become more accurate. 
However, from Fig.~\ref{fig:Cryst_cri_line_fss} we observe  that the influence of tricriticality is too big, and the scaling is closer to 
the one expected to be seen at tricritical point: $\Delta_1\sim N^{2/3}$ versus $\Delta_1\sim N^{1/2}$ for second-order phase transitions.

\begin{figure}[h!]
    \centering
    \includegraphics[width=0.7\linewidth]{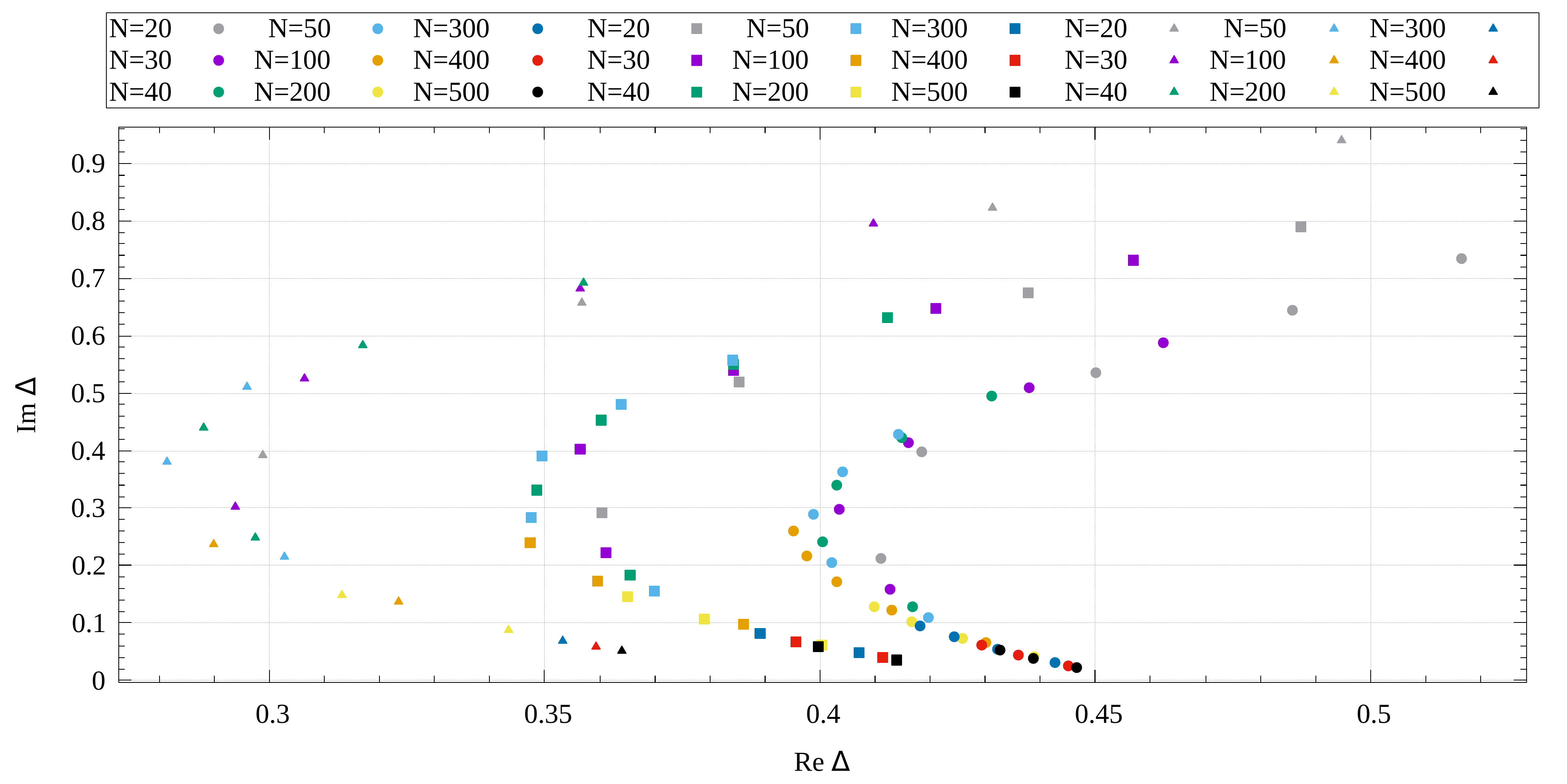}
    \caption{Crystal field zeros for different system sizes at different points along the critical line ($T=0.5, 0.4, 0.35$) (rhombes, squares, triangles). We see the zeros bend away from the real axis for smaller systems so that the angles cannot be accessed properly. }
    \label{fig:Cryst_plot_angles}
\end{figure}

However, we could not recreate the impact angle analysis along the critical line, due to the fact that the crystal field zeros bend, as depicted in Fig.~\ref{fig:Cryst_plot_angles}. The line between the first two zeros often points in the opposite direction of the expected angle. 

 \subsubsection{Near the tricritical point: $\Delta_t=2/3\ln 2$, $T=T_t=1/3$, $H_{t}=0$ }\label{results_exact_crystal_Tricritical}

\begin{figure}[h!]
    \centering
    \includegraphics[width=0.7\linewidth]{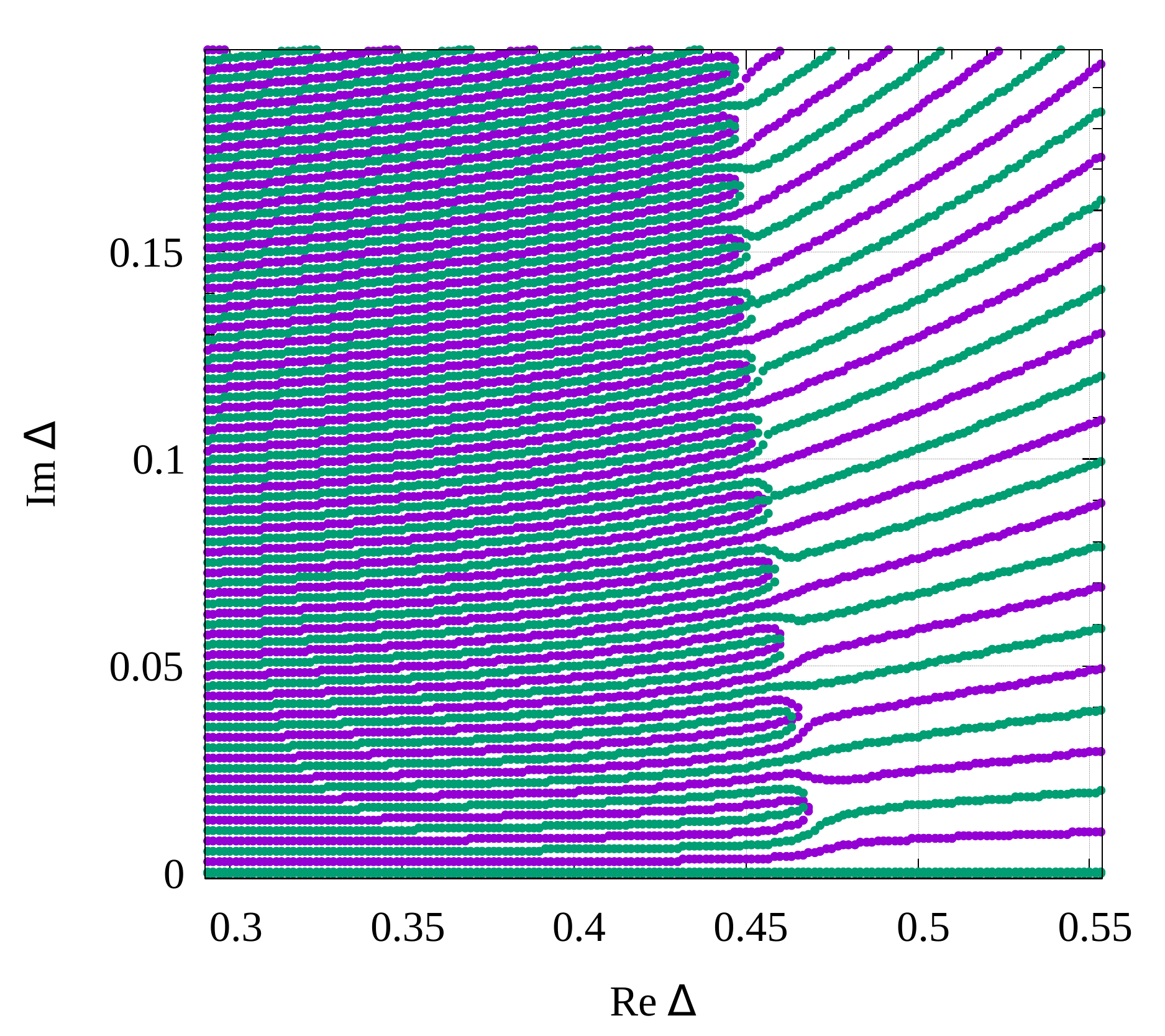}
    \caption{Solutions of equations ${\rm Re}\, Z=0$ and ${\rm Im}\, Z=0$ in complex crystal field plane for $N=200$ at $T_t$. The crystal field zeros are clearly visible. }
    \label{fig:Cryst_tri}
\end{figure}

\begin{figure}[h!]
    \centering
    \includegraphics[width=0.7\linewidth]{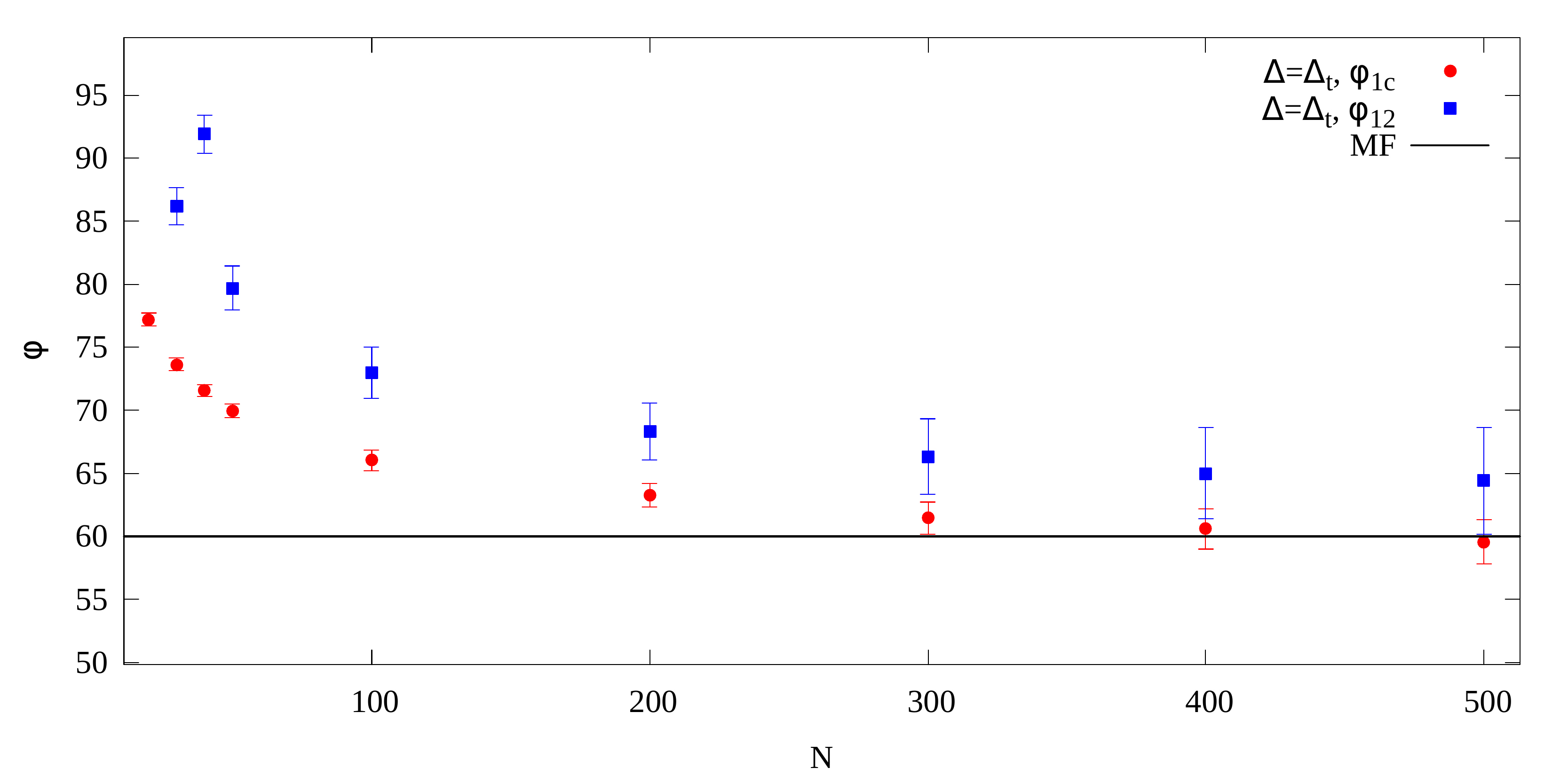}
    \caption{Angles $\varphi_{1c}$ and $\varphi_{12}$ 
    between the crystal field zeros and real crystal field axis for different system sizes at $T_t$. Both angles come within an error bar to the expected value $\varphi_{MF}^{tri}=60^\circ$ before graphs reach size $N=500$.}
    \label{fig:Cryst_plot_angles_2}
\end{figure}

Finally, at the tricritical point, crystal field zeros are distinguished in the most accurate way (Fig.~\ref{fig:Cryst_tri}).  We can also look at the impact angles for both $\varphi_{1c}$ and $\varphi_{12}$. In Fig.~\ref{fig:Cryst_plot_angles_2} both of the angles eventually converge within error bars to the theoretical mean-field value $\pi/3$.

\subsection{Lee-Yang zeros}\label{results_exact_LY}

\begin{figure}
    \centering
    \includegraphics[width=0.8\linewidth]{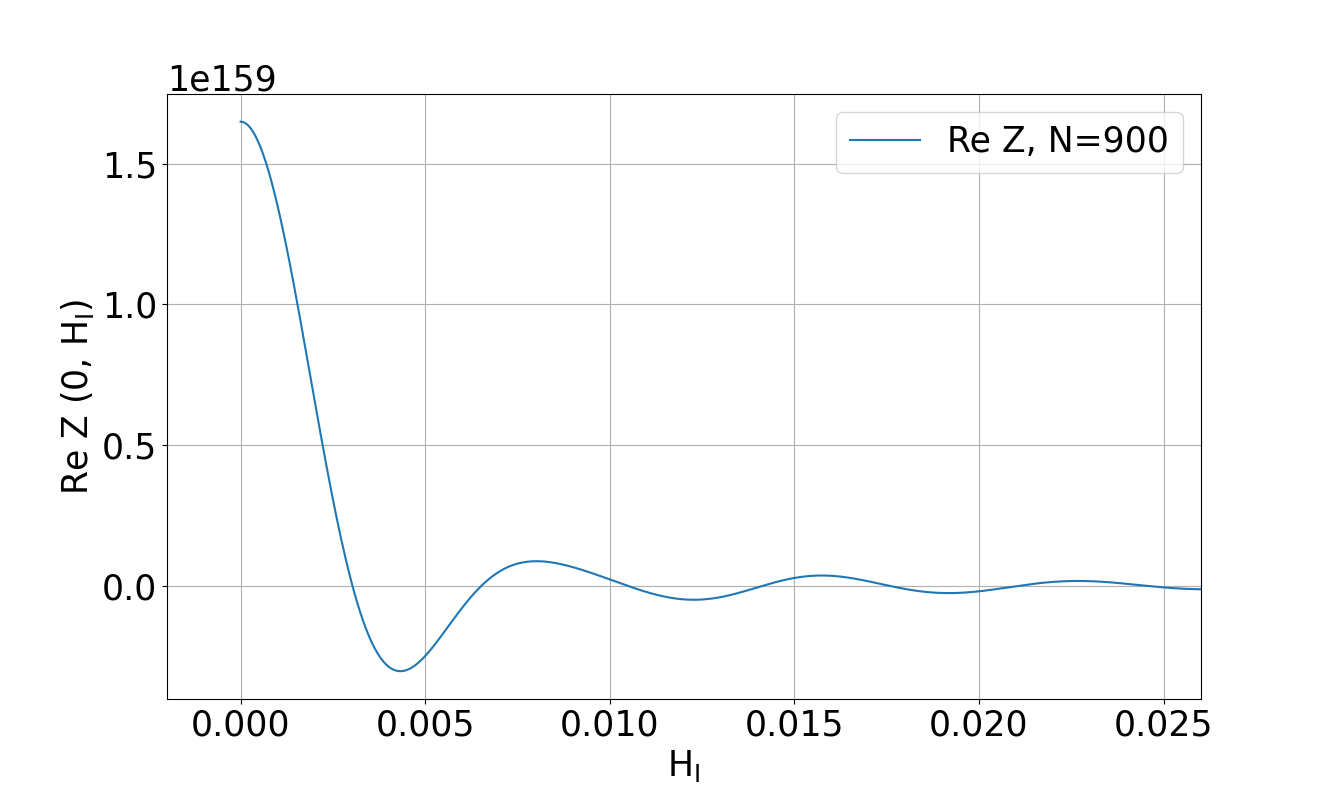}
    \caption{Function ${\rm Re}\, Z (H_R=0,H_I)$ at $N=900$. Solutions of the equation ${\rm Re}\, Z (H_R=0,H_I)=0$
    are the Lee-Yang zeros. The first solution $H_1$ is the first zero whose scaling we will follow through. }
    \label{fig:LY_rez}
\end{figure}

In this subsection we check whether the scaling form (\ref{I.10}) holds for the Lee-Yang zeros of
the exact partition function at ${\rm Re}\, H=0$ and $H=i\, {\rm Im}\, H$:
\begin{equation}
Z_N (T,\Delta,H) =
\int_{-\infty}^{+\infty} \exp\Big \{ \frac{-Nx^2}{2T}+N \ln(1+2e^{-(\frac{1}{2N}+\Delta)/T}
\cosh[(x+iH)/T])\Big \} dx \, .
\end{equation}
Since according to the Lee-Yang theorem \cite{lee_statistical_1952,yang_statistical_1952} all zeros in complex magnetic field are located along the imaginary axis, we only need to solve 
the equation ${\rm Re}\, Z = 0$. Its roots are the Lee-Yang zeros, cf. Fig.~\ref{fig:LY_rez}.

\subsubsection{Scaling exponent analysis for Lee-Yang zeros}\label{results_exact_LY_criticalpoint}

\begin{figure}[h!]
    \centering
    \includegraphics[width=0.7\linewidth]{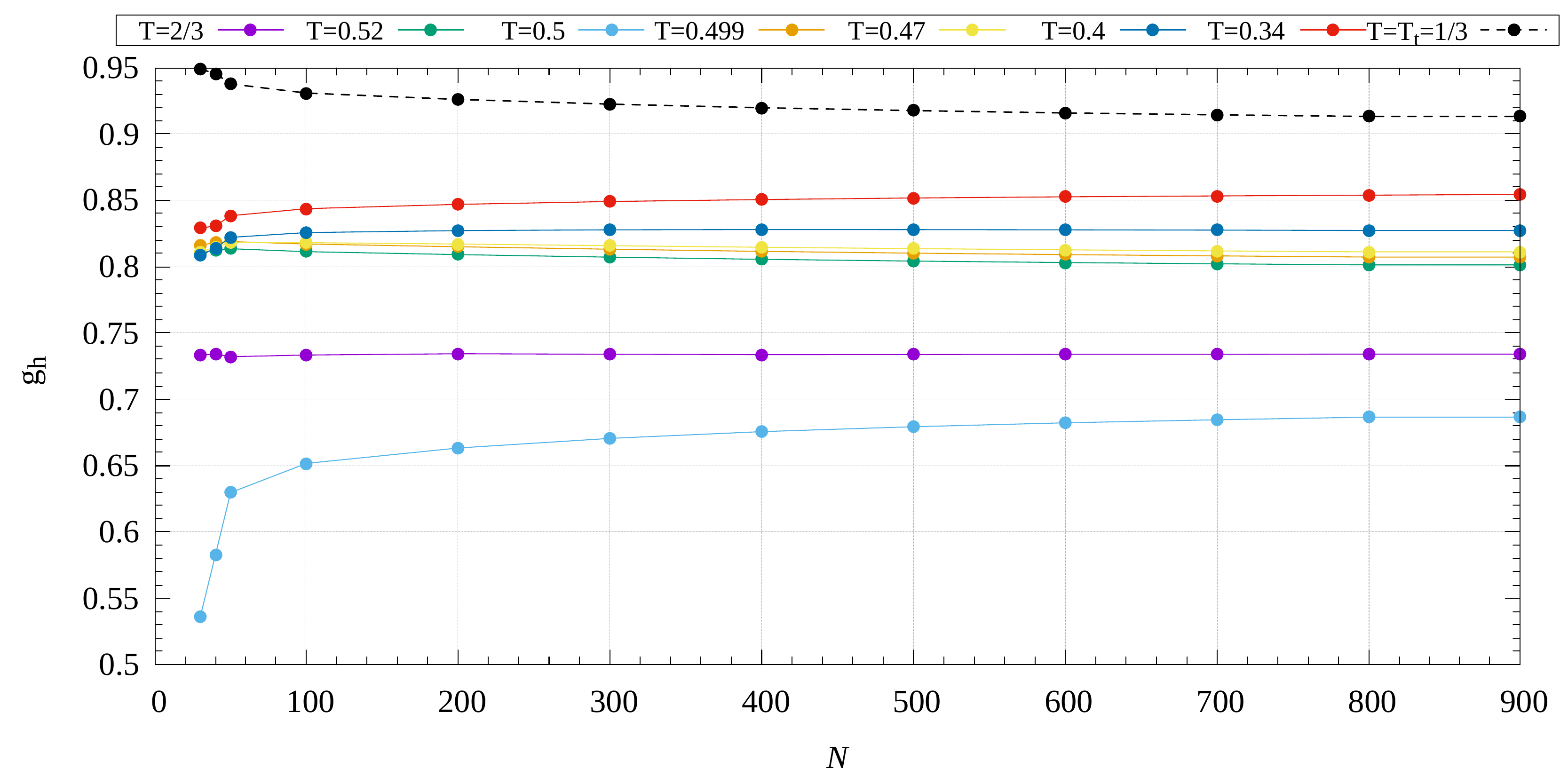}
    \caption{Scaling exponent $g_h$ that governs the FSS of the first Lee-Yang zeros $H_1$ at different points along the critical line, as well as at thee tricritical line. The approach of taking into account all of the previous graph sizes up to size $N$ when fitting the FSS is used here. Starting with the "Ising" critical point (purple line) we expect the exponents $g_h$ to converge to a mean-field value $g_h=3/4$ as the size $N$ increases. The tricritical point (black line) is expected to converge to $g_h=5/6$. The points on the critical line that are taken at some fixed non-zero crystal field are skewed towards the tricritical behaviour.}
    \label{fig:LY_yh_all}
\end{figure}

We are interested in the finite-size scaling of the first Lee-Yang zero $H_1$ along the critical line: $H\sim 0$, $T>1/3$, $\Delta\lesssim 2/3 \ln 2$ and at the tricritical point: $H= 0$, $T=1/3$, $\Delta= 2/3 \ln 2$. To this end, let us follow how the scaling exponent $g_h$ changes with the graph size $N$ for different points along the critical line. The simplest approach is to repeat the fitting process for each new size $N$ adding the point to the previous set of points (see Fig.\ref{fig:LY_yh_all}). For the infinite system size, at any point along the critical line the scaling should be governed by the same exponent
$H_1\sim N^{-g_h}=N^{-3/4}$, while at the tricritical point the tricritical scaling is $H_1\sim N^{-5/6}$. We observe that for all graph sizes under consideration
only the finite-size scaling at the "Ising point" $\Delta = 0, T = 2/3$ is governed by the proper scaling exponent, while the rest of the points along the critical line are skewed towards the tricritical scaling. The tricritical scaling itself is still far from the asymptotic exponent. Of special interest is the point on the critical line that corresponds to parameters $T=1/2, \, \Delta = \ln2 /2$. It seems to be the only critical line point that improves its scaling towards the predicted asymptotic behaviour. We will discuss the significance of this point later.

\begin{figure}[h!]
 	\centering
 	\includegraphics[width=0.7\linewidth]{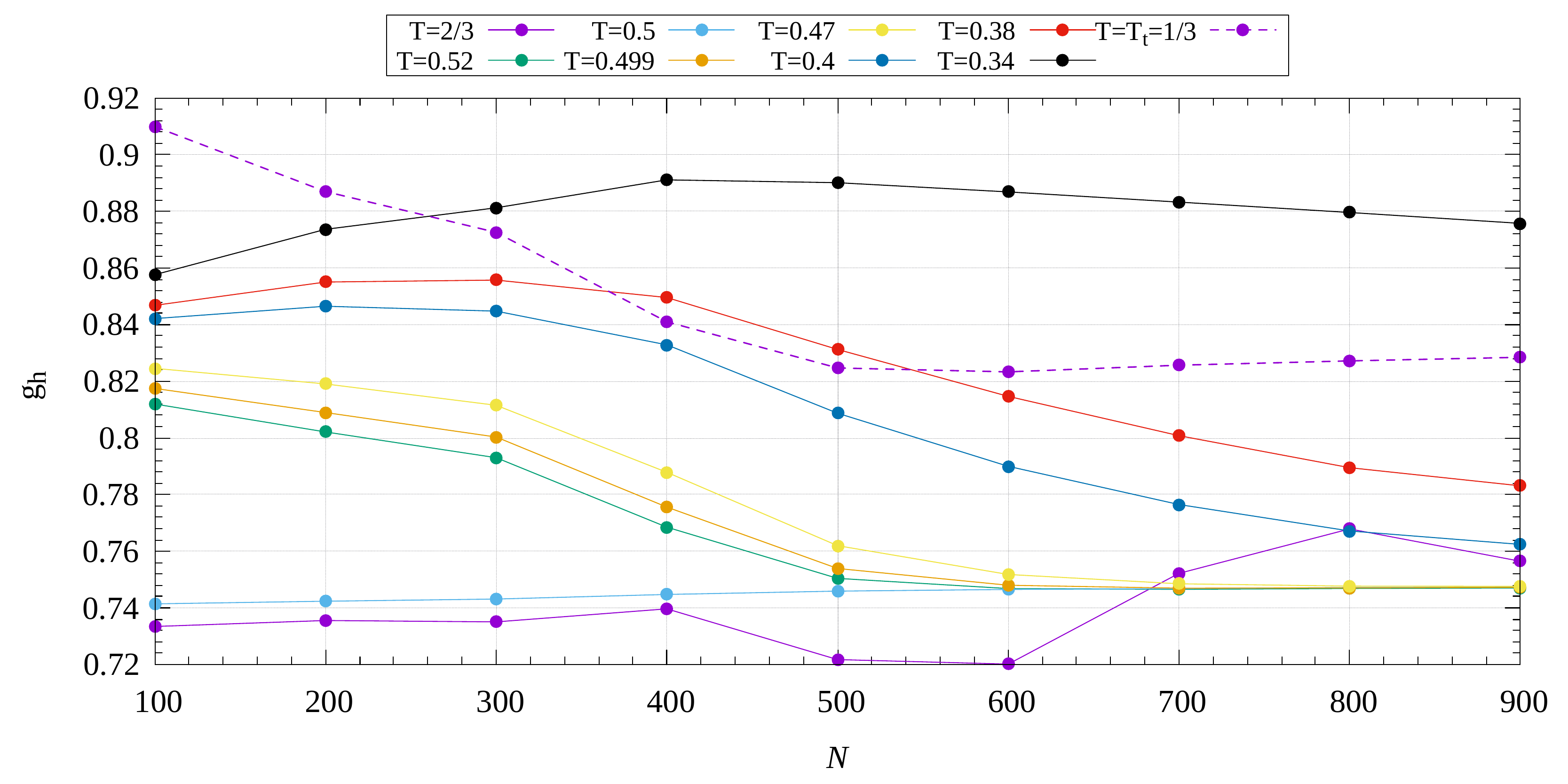}
 	\caption{Scaling exponent $g_h$ that governs the FSS of the first Lee-Yang zeros $H_1$ at different points along the critical line, as well as at the tricritical line. The approach of taking into account only five largest graph sizes up to size $N$ when fitting the FSS is used here. Starting with the "Ising" critical point (bottom purple line) the points along the critical line exhibit the scaling behaviour slightly skewed towards the tricriticality, but finally converge towards the critical value $g_h=3/4$. The tricritical point scaling converges to its own distinct value $g_h=5/6$ within the scope of our computing abilities.}
 	\label{fig:LY_yh_five}
 \end{figure}

Another approach, seeing that in most cases small system sizes corrupt the precision, is to take into account only the largest sizes (a notable counterexample is in Ref.~\cite{moueddene_critical_2024}). We do this using the five points up to the size $N$ in Fig. \ref{fig:LY_yh_five}. Here, the scaling exponent $g_h$ seems to fluctuate around its asymptotic value with increasing system size $N$. However, all other points along the critical line express much clearer direction of convergence towards $g_h=0.75$. Even though for the point that is very close to the tricritical point (corresponding to $T=0.34$ versus $T_t=1/3$) for our graph sizes the scaling exponent is even larger than that of the tricritical point ($g_h=0.8(3)$), the potential asymptotic can be established. Again, the special point $T=1/2, \Delta = \ln2 /2$ exhibits a near-perfect scaling behaviour, approaching the critical mean-field value almost from the smallest sizes.

Interestingly, the point $T=1/2, \Delta = \ln2 /2$ lies exactly on the line $\Delta=T\ln2$, described in section 
\ref{sec_model}. This is the line of a so-called pseudo-transition, that shows the transition in an unordered phase. Below this line, there are more spins with a non-zero value, while above it there are more $S=0$ spins. In both cases the mean magnetisation of the system is zero, so this transition is invisible. It is, however, an additional interesting point in the phase landscape of the Blume-Capel model. 
 
\section{Conclusions}\label{sec_conclusions}

This study explored the finite-size behaviour of the Blume-Capel model on a complete graph by analyzing the partition function zeros in the complex parameters plane: magnetic field (Lee-Yang zeros), temperature (Fisher zeros) and crystal field (Crystal Field zeros) planes. These zeros were used to characterize the nature of phase transitions and critical properties of the system using the universal relations for zeros coordinates.  We conducted a comprehensive analysis of the exact integral representation of the partition function in different critical regimes. 

The results demonstrated that Fisher zeros near the critical point are well-defined for small system sizes, although the expected scaling behaviour $t\sim N^{-1/2}$ 
was not fully reached ($t\sim N^{-0.44}$). 
The equivalence between Fisher and crystal field zeros was confirmed analytically. However, crystal field zeros at the critical point showed poor behaviour in their scaling with $\Delta_1 \sim N^{0.69}$ which is much closer the tricritical behaviour. However, the angles of Crystal field zeros condensation near the tricritical point converge with its theoretical mean-field value. 
The Lee-Yang circle theorem was also validated for zeros near the critical point, the line and the tricritical point, further supporting theoretical predictions. For Lee-Yang zeros, all zeros were found to lie on the imaginary axis, confirming the validity of the Lee-Yang circle theorem and the presence of a second-order phase transition. Scaling with the system size matched theoretical predictions, with scaling exponents $g_h \sim 3/4$ at the critical point and $g_h \sim 5/6$ at the tricritical point.

Overall, our work not only provides a deeper understanding of the Blume-Capel model on finite-size systems but also highlights the effectiveness of partition function zeros analysis for studying critical phenomena. These methods offer valuable insights for further exploration of complex systems and phase transitions.

\section*{Acknowledgments}
This paper is dedicated to Prof. Vadym Loktev to celebrate his 80th birthday. We thank Prof. Larissa Brizhik and 
Prof. Olena Gomonay, the guest editors, for the invitation to contribute to  the special issue of the 
\textit{Low Temperature Physics} dedicated to this event. We thank Andy Manapany and Le\"ila Moueddene
for many useful comments on the work presented in this paper during our regular ${\mathbb L}^4$ meetings.
M.K. and Yu.H thank for the support by the National Research Foundation
of Ukraine, Project 2023.03/0099 ``Criticality of complex systems: fundamental aspects and applications''. 
The authors thank the Ukrainian Army for the possibility to perform this research work. 
 
\bibliographystyle{unsrt}
\bibliography{Blume-Capel}

\end{document}